%% file: weak.tex
\documentclass[12pt]{article}
\pdfoutput=1 

\title{\MakeUppercase{Theory of Weak Identification in Semiparametric Models}}

\author{Tetsuya Kaji}

\date{\normalsize\today}

\usepackage[margin=1.2in]{geometry}
\usepackage[onehalfspacing,nodisplayskipstretch]{setspace} 

\usepackage{amsthm,amsmath,amssymb,calc}
\usepackage{thmtools}
\usepackage{mathrsfs}
\usepackage{mathtools}
\usepackage{bm}
\usepackage{bbm}

\usepackage{titling}
\usepackage[tiny,center]{titlesec}
\usepackage{chngcntr}
\usepackage{appendix}

\usepackage[authoryear]{natbib}
\RequirePackage[colorlinks,citecolor=blue,urlcolor=blue,breaklinks=true]{hyperref}
\usepackage{breakcites}
\RequirePackage{cleveref}
\usepackage{hyphenat} 

\usepackage{chapterbib} 
\usepackage{enumerate}
\usepackage{rotating}
\usepackage{xcolor}

\usepackage{graphicx,epstopdf,epsfig}
\usepackage{caption,subcaption}
\usepackage{pgfplots}

\pretitle{\vspace{-5em}\begin{center}\fontsize{13}{15}\selectfont}
\posttitle{\par\end{center}}
\preauthor{\vspace{-1em}\begin{center}
	\normalsize \scshape \lineskip 0.5em%
	\begin{tabular}[t]{c}}
\postauthor{\end{tabular}\par\end{center}}
\predate{\begin{center}\large}
\postdate{\par\end{center}\vspace{-1em}}

\makeatletter
\newcommand\new@setfontsize[3]{%
    \ifx \protect \@typeset@protect \let \@currsize #1\fi \fontsize {#2}{#3}\selectfont
}
\let\orig@setfontsize\@setfontsize
\let\orig@cases\cases
\let\endorig@cases\endcases

\renewenvironment{cases}{%
    \let\@setfontsize\new@setfontsize
    \setstretch{\setspace@singlespace}%
    \let\setfontsize\orig@setfontsize
    \orig@cases
}{%
    \endorig@cases
}
\makeatother

\titleformat{\section}{\normalfont\centering}{\thesection}{1em}{\MakeUppercase}
\titleformat*{\subsection}{\itshape\centering}

\theoremstyle{plain}
\newtheorem{thm}{Theorem}
\newtheorem*{thm*}{Theorem}
\newtheorem{prop}[thm]{Proposition}
\newtheorem{lem}[thm]{Lemma}

\theoremstyle{definition}
\newtheorem*{defn}{Definition}
\newtheorem{exa}{Example}
\newtheorem*{exa*}{Example}

\theoremstyle{remark}
\newtheorem*{rem}{Remark}

\renewcommand\thmcontinues[1]{continued}

\crefname{thm}{Theorem}{Theorems}
\crefname{prop}{Proposition}{Propositions}
\crefname{lem}{Lemma}{Lemmas}
\crefname{coro}{Corollary}{Corollaries}
\crefname{add}{Addendum}{Addendums}
\crefname{alg}{Algorithm}{Algorithms}
\crefname{proc}{Procedure}{Procedures}
\crefname{exe}{Exercise}{Exercises}
\crefname{exa}{Example}{Examples}
\crefname{prob}{Problem}{Problems}
\crefname{section}{Section}{Sections}
\crefname{subsection}{Section}{Sections}
\crefname{appendix}{Appendix}{Appendices}

\crefrangelabelformat{exa}{#3#1#4--#5\crefstripprefix{#1}{#2}#6}

\DeclareMathOperator{\Var}{Var}

\DeclareMathOperator{\tr}{tr}

\DeclareMathOperator{\Spn}{Span}
\DeclareMathOperator{\col}{col}

\DeclareMathOperator{\vect}{vec}
\DeclareMathOperator*{\conv}{\mathchoice{%
	\,\longrightarrow\,}{
	\rightarrow}{
	\rightarrow}{
	\rightarrow}
}

\def\ci{\perp\!\!\!\!\perp}			

\makeatletter
\renewcommand*\env@matrix[1][*\c@MaxMatrixCols c]{%
  \hskip -\arraycolsep
  \let\@ifnextchar\new@ifnextchar
  \array{#1}}
\makeatother

\makeatletter
\def\blfootnote{\gdef\@thefnmark{}\@footnotetext}
\makeatother

\allowdisplaybreaks[3]

\counterwithin{subsection}{section}

\begin{document}


\include{maintext}


\end{document}

%% file: maintext.tex

\maketitle

\begin{abstract}
We provide general formulation of weak identification in semiparametric models and an efficiency concept.
Weak identification occurs when a parameter is weakly regular, i.e., when it is locally homogeneous of degree zero.
When this happens, consistent or equivariant estimation is shown to be impossible.
We then show that there exists an underlying regular parameter that fully characterizes the weakly regular parameter.
While this parameter is not unique, concepts of sufficiency and minimality help pin down a desirable one.
If estimation of minimal sufficient underlying parameters is inefficient, it introduces noise in the corresponding estimation of weakly regular parameters, whence we can improve the estimators by local asymptotic Rao\hyp{}Blackwellization.
We call an estimator weakly efficient if it does not admit such improvement.
New weakly efficient estimators are presented in linear IV and nonlinear regression models.
Simulation of a linear IV model demonstrates how 2SLS and optimal IV estimators are improved.

\vspace{4pt}

\vspace{4pt}
\end{abstract}

\blfootnote{
I thank Anna Mikusheva, Isaiah Andrews, Victor Chernozhukov, Whitney Newey, Jerry Hausman, Hidehiko Ichimura, Kengo Kato, Rachael Meager, Peter Hull, and anonymous referees for helpful comments and suggestions. This work is supported by the Richard N.\ Rosett Faculty Fellowship and the Liew Family Faculty Fellowship at the University of Chicago Booth School of Business.}


\section{Introduction}

Weak identification arises in a wide range of empirical settings.
A leading example is the linear instrumental variables (IV) model in which the instruments and endogenous regressors are barely correlated.
When this happens, classical asymptotic theory is known to yield poor approximations to the behavior of familiar statistics \citep{ss1997}.
We encounter this problem in various other contexts: \citet{sw2000} analyze weak identification in generalized method of moments (GMM) models;
\citet{ac2012} in extremum estimation models; \citet{am2016} in dynamic stochastic general equilibrium (DSGE) models. 
Many estimators of weakly identified parameters exhibit inconsistency and bias, and, as a consequence, standard inference procedures such as $t$- and Wald tests may have substantially distorted sizes \citep{d1997,hp2015}.
Following these problems, a vast amount of theoretical work has been published.

The theoretical literature on weak identification is largely confined to specific estimation and inference procedures in specific models.
Many fundamental questions---such as what is the common cause of weak identification, what is a general guideline to find well-behaved statistics, and what is semiparametric efficiency in the presence of weak identification---are left unanswered.
Such exploration is essential, however, not only to facilitate unified understanding of the phenomenon but to measure performance of different procedures and develop systematic methods for estimation and inference.


This paper studies weak identification from the perspective of semiparametric theory (\Citealp[Chapter 25]{bkrw1993,v1998}).
This literature views parameters as functions defined on the probability manifold and relates their asymptotic properties to the functions' local behaviors in response to local perturbations of the probability.
While strongly identified parameters often translate to differentiable functions, weakly identified parameters emerge as functions that are discontinuous at the probability to which we asymptote.
We call such functions {\em weakly regular} in reference to differentiable functions called {\em regular}.
As an immediate consequence of this discontinuity, we derive---without reference to a specific estimation or inference procedure---that there exists neither a consistent estimator, a consistent test, nor an equivariant (hence pivotal) estimator when the parameter is weakly regular.
The local approximations of weakly regular parameters are homogeneous of degree zero and essentially nonlinear, becoming the root cause of non\hyp{}Gaussian nonpivotal asymptotic distributions witnessed throughout the literature \citep{ss1997,sw2000,ac2012}.
To circumvent the problem of nonlinearity, we explore weak regularity from the standpoint of {\em regular} parameters.

We show that every weakly regular parameter can be represented as a nonlinear function of the local parameter of some underlying regular parameter.
Finding such a parameter lets us reformulate the model in a way that it consists only of regular parameters, providing a tractable foundation on which to discuss estimation and inference.
This conforms with the repeated observation in the literature that reduction to regular parameters (usually referred to as ``reduced\hyp{}form parameters'') substantially simplifies the problems; we generalize this observation to arbitrary semiparametric models and show that there exists an underlying regular parameter for every weakly regular parameter.
However, underlying regular parameters are not unique, and statistical analyses based on different underlying parameters may yield different performances.
This gives rise to the need for criteria to choose an underlying parameter.

We introduce two desirable properties of underlying parameters.
In semiparametric models, the space of probability distributions can be much bigger than the space of the parameter of interest.
Consequently, there are many directions of perturbations that do not matter to the parameter of interest.
Intuitively, a good underlying parameter would be sensitive to all perturbations that matter {\em and} not sensitive to any perturbations that are irrelevant.
The first property, {\em sufficiency}, ensures that all perturbations that affect identification of the weakly regular parameter affect the underlying parameter; for example, reduced\hyp{}form coefficients in linear IV that miss an instrument are not sufficient.
The second property, {\em minimality}, guarantees that the underlying parameter does not include a nuisance parameter.
In short, a best underlying regular parameter is minimal and sufficient.
We show existence of minimal sufficient underlying parameters and present examples.

With these concepts, we define a notion of efficiency for weakly regular parameters.
Efficient estimation under weak identification has received little treatment
as non\hyp{}Gaussianity and nonpivotality of the asymptotic distributions render the classical efficiency concepts inapplicable in their direct forms.
Our formulation helps decompose estimation of weakly regular parameters into estimation of the minimal sufficient underlying regular parameters and their transformation.
Being regular, the underlying parameters are susceptible to the classical convolution theorem.
Moreover, if the estimators of the underlying parameters contain noise, their transformations, too, would suffer from noise.
Such noise can then be eliminated by taking expectation with respect to it.
Conceptually, this corresponds to applying the Rao\hyp{}Blackwell theorem to the local asymptotic representations of the estimators.
The resulting conditional expectation estimators are more concentrated around the same means.
We name this procedure {\em local asymptotic Rao\hyp{}Blackwellization (LAR)}.
If such improvement is impossible, we call the estimators {\em weakly efficient}.
We put the qualifier ``weakly'' as weakly efficient estimators are not unique.
We can interpret weak efficiency as a generalization of classical efficiency to accommodate non\hyp{}differentiable transformations.

We apply our results to linear IV and nonlinear regression models and present weakly efficient estimators.
In linear IV, the two\hyp{}stage least squares (2SLS) and even the optimal IV estimators are shown to be inefficient in the presence of heteroskedasticity and, under the availability of an efficient estimator of the reduced\hyp{}form coefficients, admit transformations into weakly efficient estimators.
In nonlinear regression, a simple least\hyp{}squares estimator is shown to be inefficient under heteroskedasticity, and we obtain a weakly efficient estimator when the heteroskedastic structure can be estimated.
In nonlinear GMM, the possibility of improvement depends on specificity of the model.
Simulation shows how weakly efficient estimators behave under weak and strong identification asymptotics in a linear IV model.

There is a large body of literature that studies the optimality of statistical procedures under weak identification.
To name a few, \citet{mw2015} study estimation that minimizes the weighted average risk when the asymptotic distribution of the statistics is known;
\citet{ams2006} develop optimal conditional likelihood ratio tests in linear IV models with normal homoskedastic errors;
\citet{m2011} studies efficient inference under a weak convergence assumption.

The rest of the paper is organized as follows.
\cref{sec:define} defines weak identification in semiparametric models, gives impossibility results, and introduces the notion of underlying regular parameters.
\cref{sec:minimal} introduces sufficiency and minimality of underlying regular parameters.
\cref{sec:lax} derives LAR for the estimation of weakly regular parameters, whence we define weak efficiency.
\cref{sec:sim} discusses application of LAR to heteroskedastic linear IV models and provides simulation results.
\cref{sec:conclusion} concludes. 
The Appendix contains proofs and the local\hyp{}to\hyp{}singularity linear IV model.

\section{Weak Identification in Semiparametric Models} \label{sec:define}

Weak identification is a problem that arises relative to the sample size.
As such, it is modeled by one of two ways: (1) start with a primitive (global) model and consider a drifting sequence, ``weak identification embedding'', toward identification failure \citep[e.g.,][]{ss1997}; (2) skip the primitive model and start with a finite\hyp{}sample situation equivalent to a local expansion around identification failure \citep[e.g.,][]{d1997}.
Both approahces are inherently local, although distinction may seem ambiguous in linear models where global and local settings coincide.
To establish a framework for efficiency, it is essential to understand the type of local expansions that can emerge from a primitive model.
Therefore, we develop a general theory for (1).

Suppose we observe i.i.d.\ random variables $X_1, \dots, X_n$ from the sample space $(\mathcal{X}, \mathscr{A})$.
The set of possible distributions of $X_i$ is denoted by $\mathcal{P}$ and called the {\em model}.
To obtain fruitful asymptotics around a distribution $P\in\mathcal{P}$, we consider a {\em path}
of distributions $Q_t \in \mathcal{P}$ indexed by a real number $t\in(0,1]$ that is {\em differentiable in quadratic mean (DQM)} at $P$, that is, there exists a measurable function $g : \mathcal{X} \to \mathbb{R}$ such that
\[
	\int_{\mathcal{X}} \biggl[ \frac{dQ_t^{1/2}-dP^{1/2}}{t} - \frac{1}{2} g dP^{1/2} \biggr]^2 \conv 0 \qquad \text{as} \qquad t \to 0,
\]
where the integral is understood with respect to a $\sigma$\hyp{}finite measure dominating $P$ and $Q_t$, and $dP$ and $dQ_t$ are their Radon\hyp{}Nikodym derivatives.
This convergence is denoted by $Q_t \conv^{\text{DQM}} P$, and we call $g$ the {\em (model) score} induced by the path $\{Q_t\}$.%
\footnote{Throughout the paper, dependence of $g$ on $\{Q_t\}$ will be implied by the context.}
The idea behind asymptotic approximation theory is that the path of ``alternatives" $\{Q_t\}$ that approaches $P$ at the same rate as the path of ``samples'' $\{\hat{P}_n\}$ is not deterministically distinguishable 
in the limit and hence yields an approximation that reflects finite sample uncertainty;
therefore, $t=1/\sqrt{n}$ under local asymptotic normality \Citep[Lemma 25.14]{v1998}, and in a minor abuse of notation we denote $Q_{1/\sqrt{n}}$ by $Q_n$.

We often do not consider every possible path in $\mathcal{P}$;
let $\mathscr{P}_P$ denote the set of paths we consider that tend to $P$ in DQM. Since there is little chance of misunderstanding, we denote $\{Q_t\}$ simply by $Q_t$, for example, $Q_t\in\mathscr{P}_P$; therefore, $Q_t$ can refer to the entire path $\{Q_t\}$ or an element $Q_t$ for a specific $t$, depending on the context.
The set $\dot{\mathcal{P}}_P$ of scores $g$ induced by the paths in $\mathscr{P}_P$ is called the {\em tangent set} at $P$.
It is clear that $\dot{\mathcal{P}}_P$ is a subset of zero\hyp{}mean functions in $L_2(P)$.%
\footnote{In this sense, $\dot{\mathcal{P}}$ is the set of {\em equivalence classes} of scores, to be precise.}
Depending on the structures of $\mathcal{P}$ and $\mathscr{P}_P$, the tangent set might be a linear space, a cone,%
\footnote{A subset $X$ of a linear space is called a {\em cone} if $x\in X$ implies $ax\in X$ for every $a>0$.}
or a set without much structure; we assume that $\mathcal{P}$ and $\mathscr{P}_P$ are nice enough that the induced tangent set is linear.
For this reason, we call the tangent set the {\em tangent space}.
The tangent space can be considered the local approximation of the model by a linear vector space.
Finally, a parameter $\psi : \mathcal{P} \to \mathbb{D}$ is defined as a map from the model $\mathcal{P}$ to a Banach space $\mathbb{D}$.%


If the parameter $\psi:\mathcal{P}\to\mathbb{D}$ is differentiable in a suitable sense, we may approximate the change in the parameter along a path by a linear map from the tangent space $\dot{\mathcal{P}}_P$ to $\mathbb{D}$.
Any infinitesimal perturbation of distribution $P$ then leads to a linear perturbation of the parameter $\psi$.
Such a parameter is known to behave well
and is said to be {\em regular}.
The appropriate notion of differentiability is as follows.

\begin{defn}[Regular parameter]
A parameter $\psi : \mathcal{P} \to \mathbb{D}$ is {\em regular} (or {\em differentiable}) at $P\in\mathcal{P}$ relative to $\mathscr{P}_P$ if there exists a continuous linear map $\dot{\psi}_P : \dot{\mathcal{P}}_P \to \mathbb{D}$ such that
\[
	\frac{\psi(Q_t)-\psi(P)}{t} \conv \dot{\psi}_P g \qquad \text{for every} \qquad Q_t \in \mathscr{P}_P.
\]
The derivative map $\dot{\psi}_P$ is called the {\em local parameter} of $\psi$.
Its adjoint map $\dot{\psi}_P^\ast:\mathbb{D}^\ast\to\overline{\dot{\mathcal{P}}_P}$ is called the {\em efficient influence map} of $\psi$, where $\mathbb{D}^\ast$ is the topological dual space of $\mathbb{D}$ and $\overline{\dot{\mathcal{P}}_P}$ the completion of $\dot{\mathcal{P}}_P$.%
\footnote{If there is a function $\tilde{\psi}_P:\mathcal{X}\to\mathbb{D}$ such that $\dot{\psi}_P^\ast\delta^\ast=\delta^\ast\tilde{\psi}_P$ for every $\delta^\ast\in\mathbb{D}^\ast$, it is called the {\em efficient influence function} \citep[Section 5.2]{bkrw1993}. 
}
\end{defn}

\begin{rem}
In the classical context, the tangent set ``represents" the set of paths, so regularity (differentiability) is often defined ``relative to the tangent set".
In the context of weak identification, however, the corresponding tangent set does not represent the set of paths
; therefore, we keep the original wording ``relative to the set of paths" from \Citet{v1991}.
The term ``regular" is taken from \Citet[Chapter 3.11]{vw1996}.%
\end{rem}

\subsection{Weakly Regular Parameters} \label{sec:wrp}

Now we define a weakly identified parameter.
Let $\mathcal{P}_\beta$ be a subset of $\mathcal{P}$ on which a parameter $\beta$ is uniquely defined.%
\footnote{$\mathcal{P}\setminus\mathcal{P}_\beta$ may contain distributions that we simply deem inconceivable as well as distributions that do not identify $\beta$.}
As the problem of weak identification arises when the population distribution is close to a point of identification failure, we model the situation by a path that takes values in $\mathcal{P}_\beta$ and approaches a point outside of $\mathcal{P}_\beta$.
However, not all such sequences are appropriate to consider. If the path approaches $\mathcal{P}\setminus\mathcal{P}_\beta$ too rapidly, $\beta$ may not be identified in the first\hyp{}order local expansion (tangent space) of $\mathcal{P}$.
To avoid this, we focus on scores that are associated with unique limiting values of $\beta$.
%

\begin{defn}[Pertinent tangent cone]
The tangent set $\dot{\mathcal{P}}_{P,\beta}\subset\dot{\mathcal{P}}_P$ {\em pertinent to} the submodel $\mathcal{P}_\beta$ at $P\in\mathcal{P}$, possibly $P\in\mathcal{P}\setminus\mathcal{P}_\beta$, is the set of scores $g\in\dot{\mathcal{P}}_P$ such that there exists a path in $\mathscr{P}_P$ that takes values in $\mathcal{P}_\beta$ and induces $g$ and every such path shares the same limit of $\beta(Q_t)$.
Define $\mathscr{P}_{P,\beta}$ to be the set of paths in $\mathscr{P}_P$ that take values in $\mathcal{P}_\beta$ and induce scores in $\dot{\mathcal{P}}_{P,\beta}$.
\end{defn}




Consequently, this paper does not cover faster\hyp{}than\hyp{}$\sqrt{n}$ weak identification. 
From the observation that $P$ is not in $\mathcal{P}_\beta$, we see that $\dot{\mathcal{P}}_{P,\beta}$ is only a cone.

\begin{lem} \label{lem:cone}
$\dot{\mathcal{P}}_{P,\beta}$ and $\dot{\mathcal{P}}_P\setminus\dot{\mathcal{P}}_{P,\beta}$ are cones.
\end{lem}

\begin{rem}
In classical asymptotic theory, the limit distribution $P$ is often regarded as the ``null hypothesis" and the path $Q_t$ as a drifting sequence of ``alternatives."
When it comes to weak identification, both the null and alternatives reside as paths in $\mathscr{P}_{P,\beta}$; $P$ is merely a point of reference for identification failure.
\end{rem}

If the set of paths $\mathscr{P}_P$ is much richer than $\mathscr{P}_{P,\beta}$ in a way that $\Spn\dot{\mathcal{P}}_{P,\beta}$ is a strict subset of $\dot{\mathcal{P}}_P$, then there exists a superfluously rich side of the model on which $\beta$ is not even defined. Since it is meaningless to consider such parts of the model when one's focus is on the parameter $\beta$, we assume innocuously that $\overline{\Spn}\,\dot{\mathcal{P}}_{P,\beta}=\overline{\dot{\mathcal{P}}_P}$.%
\footnote{Later on we define the underlying regular parameter on the whole of $\mathcal{P}$, so it is actually harmful to require that the parameter be regular on the unconsidered realm of the model.}

Now we define the weakly identified parameter under the name {\em weakly regular parameter}.
We henceforth shun the use of the qualifier ``weakly identified" since weak identification in the literature may not always exclude cases of in fact {\em no} identification.
In this paper, we require that weakly regular parameters are identified at every fixed $n$ in that there exists a unique value of the parameter for any given $Q_n$ along a path.
Moreover, we require that the parameters remain identified in the limit in the sense that there exists a unique limiting value of the parameter for each score $g$ in $\dot{\mathcal{P}}_{P,\beta}$.
Let $\mathbb{B}$ be another Banach space on which a weakly regular parameter will be defined.


\begin{defn}[Weakly regular parameter]
A parameter $\beta : \mathcal{P}_\beta \to \mathbb{B}$ is {\em weakly regular} at $P\in\mathcal{P}$, possibly $P\in\mathcal{P}\setminus\mathcal{P}_\beta$, relative to $\mathscr{P}_{P,\beta}$ if there exists a map $\beta_P : \dot{\mathcal{P}}_{P,\beta} \to \mathbb{B}$ that is continuous on $\dot{\mathcal{P}}_{P,\beta}$ (not necessarily on $\dot{\mathcal{P}}_P$) and homogeneous of degree zero such that
\[
	\beta(Q_t) \conv \beta_P(g) \qquad \text{for every} \qquad Q_t \in \mathscr{P}_{P,\beta}.
\]
\end{defn}

The definition says that the value to which a weakly regular parameter converges changes as we consider different paths.
Moreover, this dependence is homogeneous of degree zero, hence essentially nonlinear and discontinuous at $g=0$.
This makes consistent estimation impossible and asymptotic distribution nonstandard (\cref{sec:impossibility}).



\begin{rem}
Being a continuous map, a regular parameter is trivially weakly regular; that is, if $\psi:\mathcal{P}\to\mathbb{D}$ is regular, then $\psi(Q_t)\to\psi_P(g)$ where $\psi_P(g)\equiv\psi(P)$.
Also, if $\beta$ is a nontrivial weakly regular parameter, i.e., $\beta_P$ is nonconstant, then $\beta_P$ cannot be linear since a linear function that is homogeneous of degree zero must be identically zero.
\end{rem}

\begin{rem}
Homogeneity of $\beta_P$ is a natural consequence of dependence on $g$.
Since $\beta(Q_{kt})$ for fixed $k>0$ converges to the same limit as $\beta(Q_t)$, we have $\beta_P(kg)=\beta_P(g)$.
\end{rem}


Now we introduce examples and show how they satisfy our definition.

\begin{exa}[Linear IV] \label{exa:iv}
Consider the IV regression model:
\[
	\Biggl\{
	\begin{alignedat}{4}
		y_i&=x_i'\beta+\varepsilon_i=z_i'\pi\beta+u_i, \qquad \quad & \mathbb{E}[\varepsilon_i\mid z_i]&=0, \quad \mathbb{E}[u_i\mid z_i]=0, \\
		x_i'&=z_i'\pi+v_i', & \mathbb{E}[v_i\mid z_i]&=0,
	\end{alignedat}
\]
where $y_i$, $\varepsilon_i$, $u_i$ are scalars, $x_i$, $\beta$, $v_i$ are $d\times1$ vectors, $z_i$ is a $k\times1$ vector, $\pi$ is a $k\times d$ full column rank matrix, and $k\geq d$.
We show that $\beta$ is weakly regular under standard assumptions and the local\hyp{}to\hyp{}zero asymptotics of \citet{ss1997}.
The local\hyp{}to\hyp{}singularity asymptotics is discussed in \cref{sec:linear}.

Let $\mathcal{P}_{uvz}$ be the set of probability distributions $P_{uvz}$ on $(u,v',z)$ with second moments such that $\mathbb{E}[u\mid z]=0$, $\mathbb{E}[v\mid z]=0$, $\mathbb{E}[zz']$ and $\mathbb{E}[vv'\mid z]$ are invertible, 
$P_{uvz}$ dominated by the Lebesgue measure, and $dP_{uvz}$ differentiable almost everywhere in $(u,v')$.%
\footnote{Domination and differentiability are not necessary as long as each path is differentiable in quadratic mean \Citep[Section 7.2]{v1998}. We assume this for illustration of explicit derivation of scores.} 
The model $\mathcal{P}$ is the set of distributions $P$ on observables $(x,y,z)$ such that
\[
	dP(x,y,z)=dP_{uvz}(y-z'\gamma,x'-z'\pi,z) \ \ \text{for} \ \ P_{uvz}\in\mathcal{P}_{uvz}, \ \pi\in\mathbb{R}^{k\times d}, \ \gamma\in\mathbb{R}^{k\times1}.
\]
%
The submodel $\mathcal{P}_\beta$ is the subset of $\mathcal{P}$ with $\det(\pi'\pi)\neq0$ and $\gamma\in\col(\pi)$.
For $P\notin\mathcal{P}_\beta$ such that $\pi(P)=0$ and $\gamma(P)=0$, the set of pertinent paths $\mathscr{P}_{P,\beta}$ consists of paths of the form
$dQ_t(x,y,z) = dQ_{t,uvz}(y-z'(t\dot{\pi}_t\beta_t),x'-z'(t\dot{\pi}_t),z)$
for $Q_{t,uvz}$ in $\mathcal{P}_{uvz}$, $\dot{\pi}_t\to\dot{\pi}$, $\beta_t\to\beta$, and $\det(\dot{\pi}'\dot{\pi})\neq0$.
This can be seen by considering a path $Q_t$ toward $P$ such that $[\pi(Q_t)-0]/t\to\dot{\pi}$ and $[\gamma(Q_t)-0]/t\to\dot{\pi}\beta$.
If $\det(\dot{\pi}'\dot{\pi})=0$, then there are many paths taking values in $\mathcal{P}_\beta$ that have different limits of $\beta$.

Now, we characterize the scores and derive $\dot{\mathcal{P}}_{P,\beta}$ and $\beta_P$.
Being a path, $Q_{t,uvz}$ has its own model score $g_{uvz}$.%
\footnote{For example, for a parametric submodel $(u,v')\ci z$, $(u,v')'\sim N(0,\Sigma)$, $z\sim N(\zeta,I)$ with $[\zeta(Q_{t,uvz})-\zeta(P)]/t\to\dot{\zeta}$, $[\Sigma(Q_{t,uvz})-\Sigma(P)]/t\to\dot{\Sigma}$, we have $g_{uvz}=\frac{1}{2}[u\ v']\Sigma(P)^{-1}\dot{\Sigma}\Sigma(P)^{-1}\bigl[\begin{smallmatrix}u\\v\end{smallmatrix}\bigr]-\frac{1}{2}\tr(\Sigma(P)^{-1}\dot{\Sigma})+(z-\zeta(P))'\dot{\zeta}$.}
Note that the only essential restrictions on $Q_{t,uvz}$ are $\int uQ_{t,uvz}(du,dv',z)=0$ and $\int vQ_{t,uvz}(du,dv',z)=0$ for almost every $z$.
Therefore,
\[
	0=\frac{1}{t}\frac{\int u(Q_{t,uvz}-P_{uvz})(du,dv',z)}{\int P_{uvz}(du,dv',z)} \conv \frac{\int ug_{uvz}P_{uvz}(du,dv',z)}{\int P_{uvz}(du,dv',z)}=\mathbb{E}_P[ug_{uvz}\mid z].
\]
Similarly, $\mathbb{E}_P[vg_{uvz}\mid z]=0$. Thus, the set of model scores $\dot{\mathcal{P}}_{P,uvz}$ for $\mathcal{P}_{uvz}$ consists of zero\hyp{}mean functions in the $L_2(P_{uvz})$\hyp{}orthocomplement of the set of functions of the form $uf(z)$ and $vf(z)$.
With this, the model score for $Q_t$ is given by
\begin{multline*}
	\frac{dQ_t-dP}{tdP} = \frac{dQ_{t,uvz}(y-z't\dot{\pi}_t\beta_t,x'-z't\dot{\pi}_t,z)-dP_{uvz}(y-z't\dot{\pi}_t\beta_t,x'-z't\dot{\pi}_t,z)}{tdP} \\
	+\frac{dP_{uvz}(y-z't\dot{\pi}_t\beta_t,x'-z't\dot{\pi}_t,z)-dP_{uvz}(y,x',z)}{tdP} \\
	\conv g=g_{uvz}(y,x',z)-z'\dot{\pi}\beta\frac{\frac{\partial}{\partial u}dP_{uvz}}{dP}-z'\dot{\pi}\frac{\frac{\partial}{\partial v}dP_{uvz}}{dP}. 
\end{multline*}
Thus, $\dot{\mathcal{P}}_{P,\beta}$ is the set of $g$ of this form with $\det(\dot{\pi}'\dot{\pi})\neq0$.
By integration by parts,
\[
	\mathbb{E}_P[ug\mid z]=\frac{-\int uz'\dot{\pi}\beta\frac{\partial}{\partial u}dP_{uvz}(du,dv',z)-\int uz'\dot{\pi}\frac{\partial}{\partial v}dP_{uvz}(du,dv',z)}{\int dP_{uvz}(du,dv',z)}
	=z'\dot{\pi}\beta.
\]
Similarly, $\mathbb{E}_P[v'g\mid z]=z'\dot{\pi}$.
Therefore, the limit of $\beta_t$ is represented, e.g., by
\[
	\beta=(\mathbb{E}_P[zz']^{-1}\mathbb{E}_P[zv'g])^\to(\mathbb{E}_P[zz']^{-1}\mathbb{E}_P[zug])=\vcentcolon\beta_P(g),
\]
where $A^\to\vcentcolon=(A'A)^{-1}A'$ denotes the left inverse of $A$.
This map is continuous on $\dot{\mathcal{P}}_{P,\beta}$ and homogeneous of degree zero but nonlinear.
Thus, $\beta$ is weakly regular.
\end{exa}

\begin{exa}[Nonlinear regression] \label{exa:nonreg}
Consider the nonlinear regression model
\[
	y=\pi m(x;\beta)+\varepsilon, \qquad \quad \mathbb{E}[\varepsilon\mid x]=0,
\]
where $m$ is a known function that is continuously differentiable and Lipschitz in $\beta$.
Assume for ease of exposition that all variables and parameters are scalars.
The key identifying assumption is that $\mathbb{E}[y-\pi m(x;\beta)\mid x]=0$ uniquely at $(\pi,\beta)$.

Let $\mathcal{P}_{x\varepsilon}$ be the set of distributions of $(x,\varepsilon)$ such that $\mathbb{E}[\varepsilon\mid x]=0$, $dP_{x\varepsilon}$ is continuously differentiable in $\varepsilon$, and $m(x;b)$ is square-integrable for every $b$.
The model $\mathcal{P}$ on $(x,y)$ is induced by $dP(x,y)=dP_{x\varepsilon}(x,y-\pi m(x;\beta))$ for some $P_{x\varepsilon}\in\mathcal{P}_{x\varepsilon}$.
The submodel $\mathcal{P}_\beta$ is such that $\mathbb{E}[y-\pi m(x;\beta)\mid x]=0$ holds uniquely at $(\pi,\beta)$ (so $\pi\neq0$).
Pick $P\in\mathcal{P}$ with $\pi(P)=0$ and consider the paths that induce $\pi(Q_t)=t\dot{\pi}_t$, $\beta(Q_t)=\beta_t$ with $\dot{\pi}_t\to\dot{\pi}\neq0$, $\beta_t\to\beta$.
As in \cref{exa:iv}, paths $Q_{t,x\varepsilon}$ in $\mathcal{P}_{x\varepsilon}$ satisfy $\mathbb{E}[\varepsilon g_{x\varepsilon}\mid x]=0$.
\[
	\frac{dQ_t-dP}{tdP}=\frac{dQ_{t,x\varepsilon}(x,y-t\dot{\pi}_tm(x;\beta_t))-dP(x,y)}{tdP}\conv g=g_{x\varepsilon}-\dot{\pi}m(x;\beta)\frac{\frac{\partial}{\partial\varepsilon}dP_{x\varepsilon}}{dP}.
\]
Thus, $\dot{\mathcal{P}}_{P,\beta}$ is the set of $g$ of this form.
By integration by parts, $\mathbb{E}_P[\varepsilon g\mid x]=\dot{\pi}m(x;\beta)$.
Therefore, $(\dot{\pi}_Pg,\beta_P(g))$ can be given as the minimizer of $\mathbb{E}_P[(\mathbb{E}_P[\varepsilon g\mid x]-cm(x;b))^2]$ with respect to $(c,b)$.
This is homogeneous of degree zero and continuous for $\beta_P$; hence $\beta$ is weakly regular.
\end{exa}

\begin{exa}[Nonlinear GMM] \label{exa:ngmm}
Consider a nonlinear moment condition that identifies a parameter $(\pi,\beta)\in\mathbb{E}\times\mathbb{B}\subset\mathbb{R}^k\times\mathbb{R}^d$,
\[
	\mathbb{E}[M_i(\pi,\beta)]=m(\pi,\beta)=0
\]
for a random process $M_i$ (e.g., $Z_ih(X_i;c,b)$ for some $X_i$ and $Z_i$), indexed by $(c,b)$, and $\ell\geq k+d$.
%
Let $\mathbb{D}$ be the space of Lipschitz functions $m:\mathbb{E}\times\mathbb{B}\to\mathbb{R}^\ell$, equipped with the Sobolev\hyp{}type norm $\|m\|\vcentcolon=\|m\|_\infty+\|dm/d\pi'\|_\infty$.
Let $\mathcal{P}_M$ be the set of probability distributions $P_M$ of zero-mean stochastic processes taking values in $\mathbb{D}$.
The model $\mathcal{P}$ can be represented as the set of distributions $P$ of $M_i$ such that for a Borel $B\subset\mathbb{D}$,
$P(M_i\in B)=P_m(M_i-m\in B)$ for $m\in\mathbb{D}$.
The submodel $\mathcal{P}_\beta$ is the subset of $\mathcal{P}$ whose mean function $m$ is in the subset $\mathbb{D}_{P,\beta}$ of $\mathbb{D}$ of functions that have unique zeros.
Recall that we are interested in the paths along which $m$ vanishes in $\beta$ at rate $t$, that is,
$m_t(c,b)=m_{0,n}(c)+t\dot{m}_n(c,b)$ where $m_{0,t}(\pi_t)=0$, $m_t(\pi_t,\beta_t)=0$, $\dot{m}_t\to\dot{m}$, $m_{0,t}\to m_0$, $\frac{d}{d\pi'}m_{0,t}\to\frac{d}{d\pi'}m_0$, and $m_0(\pi)=0$.
So, we can write $Q_t\conv^{\text{DQM}}P$ in $\mathcal{P}_\beta$ using a path $Q_{t,M}\conv^{\text{DQM}}P_M$ in $\mathcal{P}_M$ as
\[
	Q_t(M_i\in B)=Q_{t,M}(M_i-m_t\in B) \quad \text{for Borel $B$},
\]
where $[m_t-m_0]/t\to\dot{m}\in\mathbb{D}_{P,\beta}$, $[\pi_t-\pi]/t\to\dot{\pi}$, and $\beta_t\to\beta$.
Note that $[m_t(\cdot,\cdot)-m_0(\cdot)]/t=\int M(\cdot,\cdot)[dQ_t-dP]/t\conv\mathbb{E}_P[M(\cdot,\cdot)g]$, so the moment function is regular.
The moment conditions imply
\begin{multline*}
	0=\frac{\mathbb{E}_{Q_t}[M(\pi_t,\beta_t)]-\mathbb{E}_P[M(\pi,\beta_t)]}{t}
	=\frac{\mathbb{E}_{Q_t}[M(\pi_t,\beta_t)]-\mathbb{E}_{Q_t}[M(\pi,\beta)]}{t}\\
	+\int M(\pi,\beta)\frac{dQ_t-dP}{t}
	\conv\frac{dm_0(\pi)}{d\pi'}\dot{\pi}+\mathbb{E}_P[M(\pi,\beta)g].
\end{multline*}
So $(\dot{\pi}_P(g),\beta_P(g))\in\mathbb{R}^k\times\mathbb{B}$ can be cast as the zero of the RHS.
If we replace $g$ by $kg$ for a scalar $k$, then $(k\dot{\pi},\beta)$ gives the corresponding zero; therefore, $\dot{\pi}_P$ is homogeneous of degree one and $\beta_P$ of degree zero.
However, if $\beta_P(g_1)\neq\beta_P(g_2)$, then we have no reason to expect $\dot{\pi}_P(g_1+g_2)$ to match $\dot{\pi}_P(g_1)+\dot{\pi}_P(g_2)$; therefore, $\pi$ is ``directionally differentiable'' but not regular.
This dovetails with the fact that the distribution of $\sqrt{n}(\hat{\pi}-\pi)$ is nonstandard \citep{sw2000}.
Nonetheless, $\beta_P$ is weakly regular.
\end{exa}

\subsection{Fundamental Impossibility} \label{sec:impossibility}

The utility of our theoretical formalism can be readily harvested in the following theorem.
It gives a formal proof to the conventional wisdom that a ``weakly identified" parameter cannot be estimated consistently or pivotally \citep[see, {\em inter alia},][]{ss1997,sw2000,ac2012}---but not as a characteristic of a specific estimation method---as a direct consequence of the characteristic of the model.%
\footnote{Consistent estimation may be possible in linear IV models if the number of weak instruments tends to infinity and some other conditions are met. In this case, the structural parameter is not weakly regular.}
This result can also be viewed as a generalized proof of nonexistence of a consistent test conjectured by \citet{hhm2011}.%
\footnote{Their setup can be translated into ours by taking $\mathbb{B}$ to be the product space for two estimators compared in the Hausman test, observing that a regular parameter is trivially weakly regular.}
Distinct but related are the impossibility results by \citet{d1997} and \citet{hp2015}; their setup is a generalization of the weak linear IV structure whereas our setup is a generalization of the weak identification phenomenon.
Indeed, \citet{d1997} shows nonexistence of bounded confidence sets (which is ``stronger'' than nonexistence of consistent estimators) while there exist weakly regular parameters that admit bounded confidence sets
;%
\footnote{A bounded weakly regular parameter trivially admits a bounded confidence set.}
\citet{hp2015} show the impossibility of unbiased estimation while there exist weakly regular parameters that admit unbiased estimation \citep{aa2017}.

\begin{thm}[Impossibility of consistent and equivariant estimation] \label{thm:impossibility}
There is no consistent sequence of estimators of a nontrivial weakly regular parameter; there is no consistent sequence of nontrivial tests of a nontrivial weakly regular parameter; there is no equivariant\hyp{}in\hyp{}law sequence of estimators of a nontrivial weakly regular parameter with a separable limit law.%
\footnote{There is no known example of nonseparable Borel measures and they are usually put aside in the standard theory of weak convergence \Citep[p.\ 24]{vw1996}, so the assumption of separability is innocuous. We hereafter treat it as general impossibility of equivariant estimation.}
\end{thm}

\begin{rem}
The first two claims are straightforward given the definition.
The third claim exploits the fact that the asymptotic distribution of $\hat{\beta}_n$ is ``continuous'' in local alternatives (a consequence of Le Cam's third lemma); since $\beta(Q_n)$ is discontinuous at $P$, $\hat{\beta}_n-\beta(Q_n)$ is necessarily discontinuous at $P$, failing to be equivariant.
\end{rem}

Impossibility of equivariant estimation implies that the asymptotic distribution of any estimator of a weakly regular parameter, when centered at the true value, is nonpivotal and not consistently estimable.
However, it does not preclude the possibility that there exist {\em test statistics} whose distributions are pivotal or consistently estimable.
In fact, almost any reasonable inference procedure would be based on statistics whose asymptotic distributions are known or estimable;
hence, the problem of estimation and the problem of inference bear quite distinct aspects when it comes to weakly regular parameters. This is in stark contrast to the classical context of regular parameters, in which efficient estimation and ``efficient'' inference are closely related to each other.%
\footnote{\Citet[Chapter 25]{v1998} states that ``[s]emiparametric testing theory has little more to offer than the comforting conclusion that tests based on efficient estimators are efficient.''}
This separation partly explains the specialty of current literature on inference problems regarding weak identification.


\subsection{Underlying Regular Parameters} \label{sec:urp}

The idea to analyze weak regularity is that in many cases there exists another parameter that is regular and whose local parameter controls the limit behavior of the weakly regular parameter.
In the literature, such a parameter is known as the ``reduced\hyp{}form parameter'' and is utilized in various robust inference procedures under weak identification.%
\footnote{On the other hand, the weakly regular parameter is often referred to as the ``structural parameter.''}
Then, the weakly regular parameter acts by itself as a transformation of the local parameter of some ``underlying" regular parameter; in other words, it is sufficient to know the value of (the local parameter of) the underlying regular parameter in order to infer the value of the weakly regular parameter around the point of identification failure.
We now formalize this idea.

\begin{defn}[Underlying regular parameter]
Let $\beta:\mathcal{P}_\beta\to\mathbb{B}$ be weakly regular at $P\in\mathcal{P}$ relative to $\mathscr{P}_{P,\beta}$. The parameter $\psi:\mathcal{P}\to\mathbb{D}$ is an {\em underlying (regular) parameter} for $\beta$ at $P$ relative to $\mathscr{P}_P$ if it is regular at $P$ relative to $\mathscr{P}_P$ and there exists a continuous map $\beta_{P,\psi}:\mathbb{D}_{P,\beta}\to\mathbb{B}$ that is homogeneous of degree zero such that
\[
	\beta(Q_t) \conv \beta_{P,\psi}(\dot{\psi}_Pg) \qquad \text{for every} \qquad Q_t \in \mathscr{P}_{P,\beta},
\]
where $\mathbb{D}_{P,\beta}\vcentcolon=\{\delta\in\mathbb{D}:\delta=\dot{\psi}_Pg \text{ for some } g\in\dot{\mathcal{P}}_{P,\beta}\}$.
\end{defn}



\begin{rem}
A global sufficient condition for an underlying regular parameter is that there exists a map from regular $\psi$ to weakly regular $\beta$ that is continuous on $\psi(\mathcal{P}_\beta)\subset\mathbb{D}$.
\end{rem}

\begin{rem}
\citet{c2016} defines the reduced\hyp{}form parameter as a function of the structural parameter.
We take the opposite route: the weakly regular parameter approaches a function of (the local parameter of) an underlying regular parameter.
\end{rem}

This definition requires that knowing the local parameter of the underlying regular parameter is enough to recover the value of the weakly regular parameter, that is, the reduction of information from knowing $g$ to knowing $\dot{\psi}_Pg$ does not impair the ability to discern $\beta$ in the limit.

As it turns out, it is straightforward to show that an underlying parameter always exists.
If we regard the root likelihood ratio $Q\mapsto dQ^{1/2}/dP^{1/2}$ as a parameter, it is trivially an underlying regular parameter for any weakly regular parameter.
However, whether there exists an underlying regular parameter that admits $\sqrt{n}$ estimation is a different matter.
For this, we need to investigate each model separately.

\begin{lem}[Existence of underlying regular parameter] \label{thm:exist}
Let $\beta:\mathcal{P}_\beta\to\mathbb{B}$ be weakly regular.
Then, there exist a Banach space $\mathbb{D}$ and an underlying regular parameter $\psi:\mathcal{P}\to\mathbb{D}$ for $\beta$.
\end{lem}

Below, we see that the natural parameters that appear in examples constitute underlying regular parameters. 

\begin{exa}[continues=exa:iv,name=Linear IV] \label{exa:iv:underlying}
Define $\psi\vcentcolon=(\gamma,\vect(\pi))$ to be the $(k+kd)\times1$ parameter of ``reduced\hyp{}form coefficients.''
We verify that $\psi$ is an underlying regular parameter for $\beta$.
Recall that $\dot{\gamma}=\mathbb{E}_P[zz']^{-1}\mathbb{E}_P[zug]$ and $\dot{\pi}=\mathbb{E}_P[zz']^{-1}\mathbb{E}_P[zv'g]$, that is, the local parameter of $\psi$ is a continuous linear functional of the score; therefore, $\psi$ is regular with $\dot{\psi}_Pg=(\dot{\gamma},\dot{\pi})$.
For $g\in\dot{\mathcal{P}}_{P,\beta}$, we have $\dot{\gamma}=\dot{\pi}\beta$ and $\beta_P(g)=\dot{\pi}^\to \dot{\gamma}$.
Therefore, $\psi$ is an underlying regular parameter for $\beta$ with $\beta_{P,\psi}(\dot{\psi})=\dot{\pi}^\to\dot{\gamma}$ defined on $\mathbb{D}_{P,\beta}=\{(\dot{\gamma},\vect(\dot{\pi}))\in\mathbb{R}^{k\times1}\times\mathbb{R}^{k\times d}:\det(\dot{\pi}'\dot{\pi})\neq0,\,\dot{\gamma}\in\col(\dot{\pi})\}$.
In fact, this underlying parameter admits the direct representation $\beta(Q_t)=\pi(Q_t)^\to\gamma(Q_t)$.

There are other choices of the underlying regular parameter.
Let $\pi_d$ and $\gamma_d$ be the first $d\times d$ submatrix and $d\times1$ subvector of $\pi$ and $\gamma$.
Then $\psi_{d}\vcentcolon=(\gamma_d,\vect(\pi_d))$ is also an underlying regular parameter since
$\beta_P(g)=\dot{\pi}_{d}^\to\dot{\gamma}_{d}$.
The submatrix and subvector can in fact be for any combinations of coefficients on $k$ instruments as long as $\det(\dot{\pi}_d'\dot{\pi}_d)\neq0$ and $\dot{\gamma}_d\in\col(\dot{\pi}_d)$.
This is to say that in overidentified linear IV models ($k>d$), there are many natural choices of underlying regular parameters.
\end{exa}

\begin{exa}[continues=exa:nonreg,name=Nonlinear regression]
Let $\mathbb{D}=\{cm(\cdot;b):c\in\mathbb{R},b\in\mathbb{B}\subset\mathbb{R}\}$ be the space of functions spanned by $cm(\cdot;b)$.
From the form of $\beta_P$, let us speculate that $\psi:\mathcal{P}\to\mathbb{D}$, $\psi(Q)\vcentcolon=\mathbb{E}_Q[y\mid x=\cdot]$, is an underlying regular parameter for $\beta$. At the point of identification failure $P$, we have $\psi(P)\equiv0$. Along the paths we consider,
\[
	\frac{\psi(Q_t)-\psi(P)}{t}\conv\dot{\psi}_Pg\vcentcolon=\mathbb{E}[\varepsilon g\mid x=\cdot]=\dot{\pi}m(\cdot;\beta),
\]
which shows regularity of $\psi$.
Next, $(\dot{\pi},\beta)$ can be cast as the minimizer of $\mathbb{E}_P[(\dot{\psi}_Pg(x)-cm(x;b))^2]$, which is homogeneous of degree zero and continuous for $\beta$ (ergo for $\beta_{P,\psi}$).
Conclude that $\psi$ is an underlying regular parameter for $\beta$.
It may seem surprising that $\pi$ is not a part of $\psi$, but it is encoded as the scaling factor of $\psi$.
\end{exa}

\begin{exa}[continues=exa:ngmm,name=Nonlinear GMM]
The moment function $m:\mathbb{E}\times\mathbb{B}\to\mathbb{R}$ is an underlying regular parameter for $\beta$.
Regularity is verified in the previous section.
The equation that defines $(\dot{\pi}_P,\beta_P)$ can be written as
\(
	\frac{dm_0(\pi)}{d\pi'}\dot{\pi}+\dot{m}(\pi,\beta)=0
\).
Thus, by taking $(\dot{\pi}_{P,m}(\dot{m}),\beta_{P,m}(\dot{m}))$ to be the zero of this (defined on the subset $\mathbb{D}_{P,\beta}\subset\mathbb{D}$ of functions with unique zeros), one sees that the moment function is an underlying regular parameter for $\beta$.
\end{exa}



\section{Minimal Sufficient Underlying Regular Parameters} \label{sec:minimal}

The underlying regular parameters are not unique, and different parameters may exhibit different properties.
This section develops two desirable properties for underlying parameters.
The motivation is analogous to classical semiparametric efficiency.
In a model that contains both a parameter of interest $\beta$ and a nuisance parameter, there is a variation of the data that is informative of $\beta$ and a variation that is not informative of $\beta$.
The classical theory extracts ``pure'' variation for $\beta$ and makes an estimator depend only thereon.
If $\beta$ is regular, such estimators share a unique efficient distribution by the virtue of differentiability.
In other words, the following two observations hold simultaneously: (1) desirable estimators depend only on pure variation; (2) desirable estimators share a unique distribution.

In the context of weakly regular $\beta$, (2) is no longer attainable due to \cref{thm:impossibility} while (1) still is.
In particular, we can make an estimator $\hat{\psi}$ depend only on pure variation for $\beta$ and then construct an estimator of $\beta$ using $\hat{\psi}$.
Such $\psi$ plays the role of extracting pure Gaussian variation relevant to $\beta$.
In the local expansion, this means that the local parameter $\dot{\psi}_P$ shares the same set of nuisance scores as $\beta_P$.
Note that these concepts need to be developed in local terms since the nuisance scores depend on the point of local expansion $P$.

\subsection{Nuisance Tangent Spaces} \label{sec:nts}

This section defines the space of nuisance scores for a weakly regular parameter.
Two points deserve attention.
First, a weakly regular parameter is not defined on some probability distributions.
Therefore, we do not want to deem a score nuisance if it affects identification of the weakly regular parameter.
Second, a weakly regular parameter is not linear in the local expansion.
This calls for a way to discuss nuisance\hyp{}ness for nonlinear maps.

%
%
%
In the classical semiparametric theory, local parameters are linearly related to the score, leading to a very nice use of the theory of linear operators \citep{bkrw1993}.
The following definition extends the key notion to nonlinear maps defined on a cone. 

\begin{defn}
Let $\mathscr{X}$ be a linear space and $\mathscr{Y}$ a set. For a map $f:A\to\mathscr{Y}$ defined on a cone $A$ in $\mathscr{X}$, define the {\em range} $R$ and {\em kernel} $N$ by
$R(f)\vcentcolon=\{y\in\mathscr{Y}:y=f(x)$ for some $x\in A\}$ and
$N(f)\vcentcolon=\{\tilde{x}\in\mathscr{X}:x\pm\tilde{x}\in A$ and $f(x\pm\tilde{x})=f(x)$ for every $x\in A\}$.%
\end{defn}

\begin{rem}
If $f$ is linear and $A=\mathscr{X}$, they reduce to the standard definitions of a range and a kernel for linear maps (\Citealp[p.\ 361]{v1998}; \citealp[p.\ 417]{bkrw1993}).
\end{rem}

Now we define the nuisance tangent space for $\beta$. 
Any score that does not affect identification or the value of $\beta$ only contains information about the path that is irrelevant to $\beta$, hence nuisance.
Since the tangent space $\dot{\mathcal{P}}_P$ is linear, we separate the space into the space spanned by nuisance scores and its orthocomplement.
That orthocomplement, by construction, only contains scores relevant to 
$\beta$.

\begin{defn}[Nuisance tangent space]
For a weakly regular parameter $\beta:\mathcal{P}_\beta\to\mathbb{B}$, call its kernel $N(\beta_P)\subset\dot{\mathcal{P}}_P$ the {\em nuisance tangent space} for $\beta$. 
Denote by $\Pi_\beta$ the projection operator onto $N(\beta_P)^\perp$ in $L_2(P)$.
\end{defn}

\begin{rem}
For a regular parameter $\psi$, the kernel of its local parameter $N(\dot{\psi}_P)$ corresponds to the tangent space for its nuisance parameter \Citep[p.\ 369]{v1998}.
\end{rem}

The definition indicates that $\tilde{g}\in N(\beta_P)$ means $g+\tilde{g}\in\dot{\mathcal{P}}_{P,\beta}$ and $\beta_P(g+\tilde{g})=\beta_P(g)$ for every $g\in\dot{\mathcal{P}}_{P,\beta}$; the first condition is the preservation of identification and the second the preservation of the value of $\beta$.
That is, the perturbation of $P$ in the direction of $\tilde{g}$ does not affect identification or distinction of $\beta$.
The flip side is that if $\tilde{g}\notin N(\beta_P)$, then there exists $g\in\dot{\mathcal{P}}_{P,\beta}$ such that either $g+\tilde{g}\notin\dot{\mathcal{P}}_{P,\beta}$ or $\beta_P(g+\tilde{g})\neq\beta_P(g)$, so we want our statistical procedures to be sensitive to these directions.

This lemma shows that the nuisance tangent space is indeed a linear space.

\begin{lem} \label{lem:sil}
(i) $N(\beta_P)$ is a linear space. 
(ii) If $P\in\mathcal{P}\setminus\mathcal{P}_\beta$, then $N(\beta_P)\subset\dot{\mathcal{P}}_P\setminus\dot{\mathcal{P}}_{P,\beta}$.
(iii) If $P\in\mathcal{P}\setminus\mathcal{P}_\beta$, then $g\in\dot{\mathcal{P}}_{P,\beta}$ implies $\Pi_\beta g\neq0$.
\end{lem}

Now we derive the nuisance tangent spaces and ``efficient scores'' in examples.

\begin{exa}[continues=exa:iv,name=Linear IV] \label{exa:iv:score}
We show that $N(\beta_P)=\dot{\mathcal{P}}_{P,uvz}$.
First, any $\tilde{g}$ that satisfies $\mathbb{E}_P[v\tilde{g}\mid z]=0$ and $\mathbb{E}_P[u\tilde{g}\mid z]=0$ is in $N(\beta_P)$ since $\mathbb{E}_P[v'(g+\tilde{g})\mid z]=\mathbb{E}_P[v'g\mid z]$ and $\mathbb{E}_P[u(g+\tilde{g})\mid z]=\mathbb{E}_P[ug\mid z]$, implying that $g$ and $g+\tilde{g}$ share the same $\dot{\pi}$ and $\dot{\gamma}$.
Therefore, $\dot{\mathcal{P}}_{P,uvz}\subset N(\beta_P)$.
Next, take $g_1,g_2\in\dot{\mathcal{P}}_{P,\beta}$ share the same $\dot{\pi}$ but different $\beta$.
This means that we have $\mathbb{E}_P[v'g_1\mid z]=\mathbb{E}_P[v'g_2\mid z]=z'\dot{\pi}$ and $\mathbb{E}_P[ug_1\mid z]=z'\dot{\pi}\beta_1\neq\mathbb{E}_P[ug_2\mid z]=z'\dot{\pi}\beta_2$.
If $\tilde{g}\in N(\beta_P)$, then $\mathbb{E}_P[v'(g_1+\tilde{g})\mid z]=z'c_1$, $\mathbb{E}_P[u(g_1+\tilde{g})\mid z]=z'c_1\beta_1$, $\mathbb{E}_P[v'(g_2+\tilde{g})\mid z]=z'c_2$, and $\mathbb{E}_P[u(g_2+\tilde{g})\mid z]=z'c_2\beta_2$ for some $c_1,c_2$ since $\tilde{g}$ does not affect $\beta$.
Deduce that $\mathbb{E}_P[v'\tilde{g}]=z'(c_1-\dot{\pi})=z'(c_2-\dot{\pi})$ and $\mathbb{E}_P[u\tilde{g}\mid z]=z'(c_1-\dot{\pi})\beta_1=z'(c_2-\dot{\pi})\beta_2$.
Since $\beta_1\neq\beta_2$, we must have $c_1=c_2=\dot{\pi}$.
In other words, $\mathbb{E}_P[v'\tilde{g}\mid z]=0$ and $\mathbb{E}_P[u\tilde{g}\mid z]=0$.
Therefore, $g\in\dot{\mathcal{P}}_{P,uvz}$.
%
Conclude that $N(\beta_P)=\dot{\mathcal{P}}_{P,uvz}$.

We note that we can write $g\in\dot{\mathcal{P}}_{P,\beta}$ as the sum of elements in $N(\beta_P)^\perp$ and $N(\beta_P)$.
As in \Citet[Example 25.28]{v1998}, $\Pi_\beta g=[\begin{matrix}z'\dot{\pi}\beta&z'\dot{\pi}\end{matrix}]\mathbb{E}_P\bigl[\begin{smallmatrix}u^2&uv'\\uv&vv'\end{smallmatrix}\bigm|z\bigr]^{-1}\bigl[\begin{smallmatrix}u\\v\end{smallmatrix}\bigr]$.%
\footnote{For unconditional moment restrictions, 
$\Pi_\beta g=\mathbb{E}_P\bigl[g\bigl(\begin{smallmatrix}u\\v\end{smallmatrix}\bigr)\otimes z\bigr]'\mathbb{E}_P\bigl[\bigl(\begin{smallmatrix}u^2&uv'\\uv&vv'\end{smallmatrix}\bigr)\otimes zz'\bigr]^{-1}\bigl(\begin{smallmatrix}u\\v\end{smallmatrix}\bigr)\otimes z=\bigl(\begin{smallmatrix}\mathbb{E}_P[zz']\dot{\pi}\beta\\\vect(\mathbb{E}_P[zz']\dot{\pi})\end{smallmatrix}\bigr)'\mathbb{E}_P\bigl[\bigl(\begin{smallmatrix}u^2&uv'\\uv&vv'\end{smallmatrix}\bigr)\otimes zz'\bigr]^{-1}\bigl(\begin{smallmatrix}u\\v\end{smallmatrix}\bigr)\otimes z$.}
\end{exa}

\begin{exa}[continues=exa:nonreg,name=Nonlinear regression]
We see that $N(\beta_P)=\dot{\mathcal{P}}_{P,x\varepsilon}$ and $\Pi_\beta g=\dot{\pi}m(x;\beta)\mathbb{E}_P[\varepsilon^2\mid x]^{-1}\varepsilon$ by the same argument as in \cref{exa:iv:score}.
An interesting observation here is that $\frac{\partial}{\partial\beta}m(x;\beta)\mathbb{E}_P[\varepsilon^2\mid x]^{-1}\varepsilon$ is in the closure of $\dot{\mathcal{P}}_P$ (but not necessarily in $\dot{\mathcal{P}}_{P,\beta}$).
This follows since the linearity of $\dot{\mathcal{P}}_P$ implies that $\bigl(m(x;\beta+t)\mathbb{E}_P[\varepsilon^2\mid x]^{-1}\varepsilon-m(x;\beta)\mathbb{E}_P[\varepsilon^2\mid x]^{-1}\varepsilon\bigr)/t$ is in $\dot{\mathcal{P}}_P$ for every $t>0$.
\end{exa}

\begin{exa}[continues=exa:ngmm,name=Nonlinear GMM]
Characterization of exact $N(\beta_P)$ and $\Pi_\beta g$ requires additional details, but it is clear that $\dot{\mathcal{P}}_{P,m}\subset N(\beta_P)$ since a score $\tilde{g}$ that satisfies $\mathbb{E}_P[M(\cdot,\cdot)\tilde{g}]=0$ does not affect the equation defining $(\dot{\pi}_P,\beta_P)$.
Moreover, if there exist scores $\tilde{g}$ such that $\mathbb{E}_P[M(\pi,\cdot)\tilde{g}]$ is a nonzero constant vector, then they change $\dot{\pi}_P$ but do not change $\beta_P$, so are in $N(\beta_P)$.
This is the case, for example, when the moment function is separable between $\pi$ and $\beta$.
If $M$ is a fully nonlinear function, on the other hand, this is not likely to hold.%
\end{exa}


\subsection{Sufficiency and Minimality of Underlying Regular Parameters} \label{sec:minsuf}


The underlying regular parameters are characterized by the span of their nuisance scores.
The first property we want in the underlying regular parameter is that it contain all relevant information about $\beta$.

\begin{defn}[Sufficiency of underlying regular parameter]
Let $\beta:\mathcal{P}_\beta\to\mathbb{B}$ be weakly regular.
An underlying regular parameter $\psi:\mathcal{P}\to\mathbb{D}$ for $\beta$ is {\em sufficient} if
$N(\dot{\psi}_P)\subset N(\beta_P)$.
\end{defn}

\begin{rem}
A global sufficient condition for sufficiency is that if $Q\in\mathcal{P}_\beta$ and $Q'\in\mathcal{P}$ satisfy $\psi(Q)=\psi(Q')$, then $Q'\in\mathcal{P}_\beta$.
In this sense, $\psi$ discerns identification of $\beta$.
\end{rem}

If $\psi$ is sufficient, then any perturbation that does not affect its local parameter does not affect $\beta$.
%
The following example shows that an underlying regular parameter need not be sufficient.

\begin{exa}[continues=exa:iv,name=Insufficiency in linear IV] \label{exa:iv:insuff}
Let $d=1$ and $k>1$ and consider the underlying regular parameter
$\psi_1(Q)=(\gamma_1,\pi_1)$
that induces $\dot{\psi}_{1P}g=(\dot{\gamma}_1,\dot{\pi}_1)$.
This parameter uses only the first instrument even though there are more available.
Therefore, $N(\dot{\psi}_{1P})$ contains elements $\tilde{g}\propto z_2\frac{\frac{\partial}{\partial v}dP_{uvz}}{dP}$ that only change the value of $\dot{\pi}_2$.
However, changing the value of $\dot{\pi}_2$ without changing $\dot{\gamma}_1$, $\dot{\pi}_1$, and $\dot{\gamma}_2$ makes $\beta$ unidentified; therefore, $g+\tilde{g}\notin\dot{\mathcal{P}}_{P,\beta}$ for $g\in\dot{\mathcal{P}}_{P,\beta}$, so $\tilde{g}\notin N(\beta_P)$.
Hence, $\psi$ is not sufficient.
\end{exa}

Not surprisingly, the set of all reduced\hyp{}form coefficients is sufficient.

\begin{exa}[continues=exa:iv,name=Sufficiency in linear IV] \label{exa:iv:suff}
The underlying parameter
$\psi(Q)=(\gamma,\vect(\pi))$
is sufficient.
To see this, let $\dot{\psi}_Pg=(\dot{\gamma},\vect(\dot{\pi}))$ and $\tilde{g}\in N(\dot{\psi}_P)$.
This means $\dot{\psi}_P(g+\tilde{g})=(\dot{\gamma},\vect(\dot{\pi}))$ for every $g\in\dot{\mathcal{P}}_P$.
Therefore, if $g\in\dot{\mathcal{P}}_{P,\beta}$, then $g+\tilde{g}\in\dot{\mathcal{P}}_{P,\beta}$ and $\beta_P(g+\tilde{g})=\beta_P(g)$, that is, $\tilde{g}\in N(\beta_P)$.
Conclude that $\psi$ is sufficient.
\end{exa}

\begin{rem}
The contrasting conclusion of Example 1 does not contradit the assumption $\overline{\Spn}\,\dot{\mathcal{P}}_{P,\beta}=\overline{\dot{\mathcal{P}}_P}$ since $(\dot{\pi}\beta,\vect(\dot{\pi}))$ for $\beta\in\mathbb{R}^d$ and nondegenerate $\dot{\pi}\in\mathbb{R}^{k\times d}$ spans the entire space for $(\dot{\gamma},\vect(\dot{\pi}))\in\mathbb{R}^{d+k\times d}$.
\end{rem}

The next property we want is that an underlying regular parameter captures only relevant information for the weakly regular parameter.

\begin{defn}[Minimality of underlying regular parameter] 
Let $\beta:\mathcal{P}_\beta\to\mathbb{B}$ be weakly regular.
An underlying regular parameter $\psi:\mathcal{P}\to\mathbb{D}$ for $\beta$ is {\em minimal} if
$N(\beta_P)\subset N(\dot{\psi}_P)$.
\end{defn}

\begin{rem}
A global sufficient condition for minimality is that there does not exist a non\hyp{}injective linear map $f:\mathbb{D}\to\mathbb{E}$ for some Banach space $\mathbb{E}$ such that $f(\psi)$ is also an underlying regular parameter for $\beta$.
\end{rem}

Minimality of $\psi$ requires the opposite of sufficiency.
For a minimal $\psi$, any perturbation that does not affect $\beta$ does not affect $\psi$.
In this sense, a minimal underlying parameter is free of nuisance parameters.

\begin{exa}[continues=exa:iv,name=Minimality in linear IV] \label{exa:iv:min}
From \cref{exa:iv:score} in the previous subsection, $N(\beta_P)=\dot{\mathcal{P}}_{P,uvz}$.
Since $N(\beta_P)$ is linear (\cref{lem:sil}), for every $g\in\dot{\mathcal{P}}_{P,\beta}$ and $\tilde{g}\in N(\beta_P)$, we have
\(
	g+\tilde{g}=(g_{uvz}+\tilde{g})-z'\dot{\pi}\beta\frac{\frac{\partial}{\partial u}dP_{uvz}}{dP}-z'\dot{\pi}\frac{\frac{\partial}{\partial v}dP_{uvz}}{dP}
\)
and $g_{uvz}+\tilde{g}\in N(\beta_P)$. Thus, $\dot{\psi}_P(g+\tilde{g})=\dot{\psi}_P(g)$ and $\tilde{g}\in N(\dot{\psi}_P)$ for $\psi=(\gamma,\vect(\pi))$.
In other words, $N(\beta_P)\subset N(\dot{\psi}_P)$, implying that $\psi$ is minimal.
The above argument applies verbatim to $\psi_1=(\gamma_1,\pi_1)$, so $\psi_1$ is minimal as well.%
\footnote{Note that $\psi=(\gamma,\vect(\pi))$ is still minimal even in the homoskedastic model.
Homoskedasticity helps simplify efficient estimation, but does not help simplify the semiparametric structure itself.}
\end{exa}

\begin{rem}
Minimal sufficiency in our definition is of a parameter, while minimal sufficiency in the context of sufficient statistics is of a statistic.
\end{rem}

This theorem ensures that a minimal sufficient underlying parameter exists.

\begin{thm}[Existence of minimal sufficient underlying regular parameter] \label{thm:minexist}
For every weakly regular parameter, there exists a minimal sufficient underlying regular parameter.
\end{thm}

Minimal sufficiency {\em per se} is not strong enough to pin down the underlying parameter uniquely.
However, minimal sufficient underlying parameters are almost equivalent to each other in terms of the efficient variation they can extract.

\begin{thm}[Characterization of minimal sufficient underlying regular parameter] \label{thm:minchara}
Let $\beta:\mathcal{P}_\beta\to\mathbb{B}$ be weakly regular and $\psi:\mathcal{P}\to\mathbb{D}$ a sufficient underlying regular parameter for $\beta$.
Then, $\psi$ is minimal if and only if for any sufficient underlying regular parameter $\phi:\mathcal{P}\to\mathbb{E}$ for $\beta$ on a Banach space $\mathbb{E}$ there exists a linear map $\tau:\mathbb{E}\to\mathbb{D}$ 
such that
\[
	\tau(\dot{\phi}_Pg)=\dot{\psi}_Pg \qquad \text{for every} \qquad g\in\dot{\mathcal{P}}_P.
\]
\end{thm}

Let us look at examples of minimal sufficient underlying parameters.


\begin{exa}[continues=exa:iv,name=Linear IV] \label{exa:iv2}
Without any prior knowledge of instrumental irrelevance, $\psi=(\gamma,\vect(\pi))$ is a minimal sufficient underlying regular parameter.
\end{exa}

\begin{exa}[continues=exa:nonreg,name=Nonlinear regression]
We show that the parameter $\psi$ is minimal and sufficient.
If $\tilde{g}\in\dot{\mathcal{P}}_P$ implies $\dot{\psi}_P(g+\tilde{g})=\dot{\psi}_Pg$, then by the formula of $\beta_{P,\psi}$, the value of $\beta_P$ does not change; hence $\psi$ is sufficient.
Minimality of $\psi$ is nontrivial; we show that $\dot{\psi}_P(g+\tilde{g})\neq\dot{\psi}_Pg$ implies $\tilde{g}\notin N(\beta_P)$.
From the formula of $\beta_{P,\psi}$, if $\beta_P(g_1+\tilde{g})=\beta_P(g_1)$ and $\dot{\psi}_P(g_1+\tilde{g})\neq\dot{\psi}_Pg_1$, then $\dot{\psi}_P(g_1+\tilde{g})$ can only be different from $\dot{\psi}_Pg_1$ in the value of $\dot{\pi}$; let them be $\tilde{\pi}_1m(\cdot;\beta_1)$ and $\dot{\pi}_1m(\cdot;\beta_1)$.
However, for another $g_2\in\dot{\mathcal{P}}_{P,\beta}$ with $\beta_P(g_2)=\beta_2\neq\beta_1$, $\dot{\psi}_P(g_2+\tilde{g})=\dot{\pi}_2m(\cdot;\beta_2)+\tilde{\pi}m(\cdot;\beta_1)$, which yields (if at all) a value of $\beta$ different from $\beta_2$. Therefore, $\tilde{g}\notin N(\beta_P)$.
\end{exa}

\begin{exa}[continues=exa:ngmm,name=Nonlinear GMM] \label{exa:ngmm2}
We show that the moment function $m$ is sufficient and, in some cases, minimal.
Recall that $(\dot{\pi}_P,\beta_P)$ is completely characterized by $\dot{m}_P$ through $\frac{dm_0(\pi)}{d\pi'}\dot{\pi}_P(g)+(\dot{m}_Pg)(\pi,\beta_P(g))=0$.
Therefore, if $g$ does not alter $\dot{m}_Pg$, $\beta_P(g)$ remains unchanged, showing sufficiency.
Or equivalently, we can see this by noting $N(\dot{m}_P)=\dot{\mathcal{P}}_{P,m}\subset N(\beta_P)$ from the previous section.
If $\dot{\mathcal{P}}_{P,m}=N(\beta_P)$ (intuitively, if the moment function is an involved nonlinear function), the moment function is minimial.
\end{exa}


Given a minimal sufficient underlying regular parameter, the problem of estimation or inference for a weakly regular parameter is translated without loss into a problem for the underlying parameter that is regular.



\section{Weak Efficiency for Weakly Regular Parameters} \label{sec:lax}

This section defines a notion of efficiency for the estimators of a weakly regular parameter.
The difficulty in defining efficiency is that their asymptotic distributions are nonstandard and nonpivotal (\cref{thm:impossibility}).
Just as a weakly regular parameter being locally a nonlinear transformation of an underlying regular parameter, an estimator of a weakly regular parameter is often a nonlinear transformation of the estimator of an underlying regular parameter.
Then, even when the estimator of the underlying parameter is Gaussian, its nonlinear transformation can in principle be anything.

A key observation is that if the estimator of the underlying parameter contains noise, its transformation also suffers from unnecessary noise.
Our idea to define efficiency lies in consideration of such noise; if the estimator of the weakly regular parameter is asymptotically an appropriate transformation of an efficient estimator of a minimal underlying regular parameter, we call it {\em weakly efficient}.
Here, the base estimator must be efficient for otherwise it is contaminated by noise, and the underlying parameter must be minimal for otherwise its transformation might be affected by the estimator of a nuisance component for $\beta$.
If the underlying parameter is also sufficient, then the weakly efficient estimator may become efficient under strong identification asymptotics.

It is helpful to draw analogy with classical efficiency on regular parameters.
Consider two {\em regular} parameters, $\nu\in\mathbb{B}$ and $\psi\in\mathbb{D}_\nu\subset\mathbb{D}$, related through a Hadamard differentiable map $\nu_\psi:\mathbb{D}_\nu\to\mathbb{B}$ by $\nu=\nu_\psi(\psi)$.
Since a differentiable map can be approximated locally by a continuous linear map, we may assume that $\nu_\psi$ is continuous linear when two parameters admit consistent estimation.
To construct an estimator of $\nu$ from an estimator of $\psi$ as $\hat{\nu}=T(\hat{\psi})$, there are two aspects to consider: (1) the efficiency of $\hat{\psi}$ and (2) the desirability of the map $T$.
Note that when $\hat{\psi}$ takes values in $\mathbb{D}_\nu$, there is little motivation to choose $T$ other than $\nu_\psi$.
Then, \Citet{v1991b} shows that if one has an efficient estimator $\hat{\psi}$ of $\psi$ that takes values in $\mathbb{D}_\nu$, the plug\hyp{}in estimator $\nu_\psi(\hat{\psi})$ is efficient for $\nu$.

If, on the other hand, $\hat{\psi}$ takes values in a bigger space $\mathbb{D}$, we need to consider an optimal choice of $T$.
Consider, for example, the strongly\hyp{}identified linear IV model with unconditional moment restrictions and overidentification, that is, $\mathbb{E}[z(y-x'\beta)]=0$ with $k>d$.
Rewriting $\mathbb{E}[zz']^{-1}\mathbb{E}[z(y-x'\beta)]=\gamma-\pi\beta=0$, we can let $\nu=\beta$, $\psi=(\gamma,\pi)$, and $\nu_\psi(\psi)=\pi^\to\gamma$.
Note that $\mathbb{B}=\mathbb{R}^d$, $\mathbb{D}=\mathbb{R}^{k\times d}\times\mathbb{R}^k$, and $\mathbb{D}_\nu$ is a subspace of $\mathbb{D}$ in which $\gamma$ is in the column space of $\pi$ and $\pi$ is of full column rank.
The OLS estimator $(\hat{\gamma},\hat{\pi})$, which is efficient under unconditional moments, takes values outside of $\mathbb{D}_\nu$ in that $\hat{\gamma}$ falls outside the column space of $\hat{\pi}$ with probability $1$.
Therefore, we have to extend $T$ that supports the bigger space $\mathbb{D}$.
This can be considered as a problem of regressing $\hat{\gamma}$ on $\hat{\pi}$ for which the generalized least squares (GLS) estimation is possible since the variance of the error term $\hat{\gamma}-\hat{\pi}\beta$ can be consistently estimated, yielding an optimally weighted GMM estimator for $\beta$.

When $\beta$ is weakly regular, \cref{thm:minexist} guarantees that one can find a minimal sufficient underlying regular parameter $\psi$ with which $\beta$ is related locally through $\beta_{P,\psi}$.
The key difference is that $\beta_{P,\psi}$ is not a linear map; it is a continuous homogeneous\hyp{}of\hyp{}degree\hyp{}zero map.
Nevertheless, we can construct an estimator of $\beta$ from an estimator of $\psi$ as $\hat{\beta}=T(\hat{\psi})$, taking into account the two aspects: (1) the effciency of $\hat{\psi}$ and (2) the desirability of $T$.
The major consequence we must face, however, is that, since the relation of $\beta$ and $\psi$ is no longer linear, there are multiple choices of $T$ that are admissible from various perspectives.
In fact, even when $\hat{\psi}$ takes values in $\mathbb{D}_{P,\beta}$, it may make sense to consider $T$ other than $\beta_{P,\beta}$, which we illustrate by an example.

In this section, we first define what it means for an estimator to be of the form $T(\hat{\psi})$.
Then, we show that this estimator admits improvement via Rao\hyp{}Blackwellization when an efficient estimator for $\psi$ is available.
This allows us to define weak efficiency of an estimator among this class of estimators.
Finally, we discuss estimators that are efficient under both strong and weak identification asymptotics.

Throughout this section, we assume that there exists a $\sqrt{n}$ efficient estimator of the minimal sufficient underlying regular parameter.
Note, however, that not all regular parameters admit $\sqrt{n}$ consistent or efficient estimators in general.

\subsection{Regular Estimators}

We focus on estimators of $\beta$ that are transformations of regular estimators of $\psi$.
First, recall the regular estimator for a regular parameter.

\begin{defn}[Regular estimator for regular parameter]
A sequence of estimators $\hat{\psi}_n$ for a regular parameter $\psi:\mathcal{P}\to\mathbb{D}$ is called {\em regular} at $P\in\mathcal{P}$ relative to $\mathscr{P}_P$ if there exists a tight Borel random element $L$ in $\mathbb{D}$ such that
\[
	\sqrt{n}(\hat{\psi}_n-\psi(Q_n)) \overset{Q_n}{\leadsto} L \qquad \text{for every} \qquad Q_n \in \mathscr{P}_P.
\]
This sequence is called {\em (semiparametric) efficient} at $P$ relative to $\mathscr{P}_P$ if it attains the distributional lower bound (denote it by $L_\psi$) of the convolution theorem.
\end{defn}

\begin{rem}
The convolution theorem states that $L=L_\psi+L_\eta$ where $L_\psi$ and $L_\eta$ are independent tight Borel random elements in $\mathbb{D}$ such that $\Pr\bigl(L_\psi\in \overline{R(\dot{\psi}_P)}\bigr)=1$ and
$\delta^\ast L_\psi \sim N \bigl( 0,\|\dot{\psi}_P^\ast\delta^\ast\|_{L_2(P)}^2 \bigr)$ for every $\delta^\ast\in\mathbb{D}^\ast$ \Citep[Theorem 2.1]{v1991b}.
This is to say, the asymptotic distribution of any regular estimator of a regular parameter is the sum of a Gaussian variable with covariance being the ``$L_2$ norm" of the efficient influence map and an independent noise. It is efficient when $L_\eta\equiv0$.
\end{rem}

\begin{rem}
If we center $\hat{\psi}_n$ at $\psi(P)$, then $\sqrt{n}(\hat{\psi}_n-\psi(P))\leadsto^{Q_n}\dot{\psi}_Pg+L$.
\end{rem}

We restrict the class of estimators of $\beta$ to ones that asymptote to functions of estimators of a minimal underlying parameter.
Considering a non\hyp{}minimal underlying parameter per se is not a problem when such $\psi$ can be estimated efficiently, but we require that the estimator of $\beta$ depends only on its minimal component asymptotically.

%

\begin{defn}[Regular estimator for weakly regular parameter]
A sequence of estimators $\hat{\beta}_n$ for a weakly regular parameter $\beta:\mathcal{P}_\beta\to\mathbb{B}$ is called {\em regular} at $P\in\mathcal{P}$ relative to $\mathscr{P}_{P,\beta}$ if there exist a minimal underlying regular parameter $\psi:\mathcal{P}\to\mathbb{D}$ for $\beta$, a regular sequence of estimators $\hat{\psi}_n$ of $\psi$, and a function $T:\mathbb{D}\to\mathbb{B}$ that is $(\dot{\psi}_Pg+L)$\hyp{}almost everywhere continuous for every $g\in\dot{\mathcal{P}}_{P,\beta}$ such that
\[
	\hat{\beta}_n=T\bigl(\sqrt{n}(\hat{\psi}_n-\psi(P))\bigr)+o_P(1) \qquad \text{for every} \qquad Q_n\in\mathscr{P}_{P,\beta}.
\]
\end{defn}

The asymptotic distribution of a regular estimator follows straightforwardly from the continuous mapping theorem.

\begin{prop} \label{thm:dist}
Let $\hat{\beta}_n=T(\sqrt{n}(\hat{\psi}_n-\psi(P)))+o_P(1)$ be a regular sequence of estimators for a weakly regular parameter $\beta:\mathcal{P}_\beta\to\mathbb{B}$.
Then,
\[
	\hat{\beta}_n\overset{Q_n}{\leadsto}T(\dot{\psi}_Pg+L).
\]
\end{prop}

\begin{exa}[continues=exa:iv,name=Linear IV] \label{exa:iv:regular}
We verify regularity of 2SLS, optimal IV, GMM, limited information maximum likelihood (LIML) , continuously updating GMM (CUE), Fuller, and unbiased \citep{aa2017} estimators.

Observe that the reduced\hyp{}form coefficients $(\gamma,\pi)$ are regular and the 2SLS can be written as a function of their estimators $\hat{\pi}_n=(Z'Z)^{-1}Z'X$ and $\hat{\gamma}_n=(Z'Z)^{-1}Z'Y$:
\[
	\hat{\beta}_{\text{2SLS}} = (\hat{\pi}_n'(Z'Z)\hat{\pi}_n)^{-1}\hat{\pi}_n'(Z'Z)\hat{\gamma}_n
	= (\sqrt{n}\hat{\pi}_n'\mathbb{E}[zz']\sqrt{n}\hat{\pi}_n)^{-1}\sqrt{n}\hat{\pi}_n'\mathbb{E}[zz']\sqrt{n}\hat{\gamma}_n + o_P(1).
\]
The residual is $o_P(1)$ since $(Z'Z)/n$ converges to $\mathbb{E}[zz']$ in probability under every path.
Since $\hat{\pi}_n$ is of full column rank with probability $1$, $T(\gamma,\pi)=(\pi'\mathbb{E}[zz']\pi)^{-1}\pi'\mathbb{E}[zz']\gamma$ is continuous $(\hat{\gamma}_n,\hat{\pi}_n)$\hyp{}almost everywhere.

Under the conditional moment restrictions
$\mathbb{E}\bigl[\begin{smallmatrix}y-x'\beta\\x-\pi'z\end{smallmatrix}\bigm|z\bigr]=0$,
the optimal IV is a $(d+dk)\times(1+d)$ matrix of the form
$C\bigl(\begin{smallmatrix}\pi'z\\&I_d\otimes z\end{smallmatrix}\bigr)\mathbb{E}\bigl[\begin{smallmatrix}\varepsilon^2&\varepsilon v'\\\varepsilon v&vv'\end{smallmatrix}\bigm|z\bigr]^{-1}$
for any $(d+dk)\times(d+dk)$ full\hyp{}rank matrix $C$ \citep{n1993}.
In fact, we can ignore $C$ and $\pi$ and use the $(k+dk)\times(k+dk)$ matrix
\(
	A(z)\vcentcolon=(I_{1+d}\otimes z)\mathbb{E}\bigl[\begin{smallmatrix}\varepsilon^2&\varepsilon v'\\\varepsilon v&vv'\end{smallmatrix}\bigm|z\bigr]^{-1}
\).
Note that $A(z)$ cannot be consistently estimated because of $\varepsilon$.
The optimal IV estimator $(\hat{\beta}_n,\hat{\pi}_n)$ minimizes
$\mathbb{E}_n\bigl[\hat{A}(z)\bigl(\begin{smallmatrix}y-x'\hat{\beta}_n\\x-\hat{\pi}_n'z\end{smallmatrix}\bigr)\bigr]'\mathbb{E}_n\bigl[\hat{A}(z)\bigl(\begin{smallmatrix}y-x'\hat{\beta}_n\\x-\hat{\pi}_n'z\end{smallmatrix}\bigr)\bigr]$, that is,
\begin{multline*}
\begin{aligned}
	\biggl[\begin{matrix}\hat{\beta}_n\\\vect(\hat{\pi}_n)\end{matrix}\biggr]&=\mathbb{E}_n\biggl[\hat{A}(z)\biggl(\begin{matrix}x'\\&I_d\otimes z'\end{matrix}\biggr)\biggr]^\to\mathbb{E}_n\biggl[\hat{A}(z)\biggl(\begin{matrix}y\\x\end{matrix}\biggr)\biggr]\\
	&=\biggl(\mathbb{E}_n[\hat{A}(z)(I_{1+d}\otimes z')]^{-1}\mathbb{E}_n\biggl[\hat{A}(z)\biggl(\begin{matrix}x'\\&I_d\otimes z'\end{matrix}\biggr)\biggr]\biggr)^\to
\end{aligned}\\
	\mathbb{E}_n[\hat{A}(z)(I_{1+d}\otimes z')]^{-1}\mathbb{E}_n\biggl[\hat{A}(z)\biggl(\begin{matrix}y\\x\end{matrix}\biggr)\biggr].
\end{multline*}
Then, we can think of $\mathbb{E}_n[\hat{A}(z)(I_{1+d}\otimes z')]^{-1}\mathbb{E}_n\bigl[\hat{A}(z)\bigl(\begin{smallmatrix}x'\\&I_d\otimes z'\end{smallmatrix}\bigr)\bigr]$ as a weighted least squares (WLS) estimator of $\bigl[\begin{smallmatrix}\pi\\&I_{dk}\end{smallmatrix}\bigr]$ and $\mathbb{E}_n[\hat{A}(z)(I_{1+d}\otimes z')]^{-1}\mathbb{E}_n\bigl[\hat{A}(z)\bigl(\begin{smallmatrix}y\\x\end{smallmatrix}\bigr)\bigr]$ as a WLS estimator of $\bigl[\begin{smallmatrix}\gamma\\\vect(\pi)\end{smallmatrix}\bigr]$.
Thus, if $\hat{A}$ is a continuous function of an estimator of the reduced\hyp{}form coefficients, the optimal IV estimator is regular.

For GMM, let $W$ be the weighting matrix.
The GMM estimator $\hat{\beta}_{\text{GMM}}$ solves
$\min_b[Z'(Y-Xb)]'W[Z'(Y-Xb)]$.
Write the objective function as
\[
	\sqrt{n}(\hat{\gamma}_n-\hat{\pi}_nb)'Z'ZWZ'Z\sqrt{n}(\hat{\gamma}_n-\hat{\pi}_nb).
\]
The optimal $W$ under unconditional moment restrictions is
$\mathbb{E}[(y-x'\beta)^2zz']^{-1}$,
and its feasible version is $\hat{W}_{\text{2SGMM}}=\mathbb{E}_n[(y-x'\hat{\beta}_{\text{2SLS}})^2zz']^{-1}$.
The expectation involved in $W$ (other than the 2SLS estimator) can be consistently estimated.
Being a function of the reduced\hyp{}form OLS and 2SLS, the two\hyp{}step GMM is regular.

LIML estimates $W$ assuming homoskedasticity, i.e., $\hat{W}_{\text{LIML}}(b)=n(Z'Z)^{-1}/\hat{\sigma}^2(b)$ where $\hat{\sigma}^2(b)=\mathbb{E}_n[(y-x'b)^2]$. 
Since the second and cross moments of $y$ and $x$ can be consistently estimated, LIML is asymptotically only a function of the OLS estimators of the reduced\hyp{}form coefficients.
Similarly, the continuously updating GMM is regular as it uses $\hat{W}_{\text{CUE}}(b)=\mathbb{E}_n[(y-x'b)^2zz']^{-1}$, which, again, admits consistent estimation.

For Fuller, let $P\vcentcolon=Z(Z'Z)^{-1}Z'$.
For a constant $C$, let $\hat{P}_{\text{Fuller}}\vcentcolon=P+(C/n)(I-P)$.
The Fuller estimator is then given by
\begin{align*}
	\hat{\beta}_{\text{Fuller}}&=(X'\hat{P}_{\text{Fuller}}X)^{-1}(X'\hat{P}_{\text{Fuller}}Y)\\
	&=\bigl(C\mathbb{E}[xx']+\sqrt{n}\hat{\pi}_n'\mathbb{E}[zz']\sqrt{n}\hat{\pi}_n\bigr)^{-1}\bigl(C\mathbb{E}[xy]+\sqrt{n}\hat{\pi}_n'\mathbb{E}[zz']\sqrt{n}\hat{\gamma}_n\bigr)+o_P(1).
\end{align*}
Thus, Fuller yields a ``weighted combination" of OLS ($C=\infty$) and 2SLS ($C=0$).

Finally, the unbiased estimator is regular.
For simplicity, let $d=1$ and $k=1$ and assume that we know $\pi>0$.
Also denote the asymptotic variance of $\sqrt{n}(\hat{\gamma}_n,\hat{\pi}_n)$ by $\bigl(\begin{smallmatrix}\sigma_\gamma^2&\sigma_{\gamma\pi}\\\sigma_{\gamma\pi}&\sigma_\pi^2\end{smallmatrix}\bigr)$.
Then,
\[
	\hat{\beta}_{\text{unbiased}}=\frac{1-\Phi(\sqrt{n}\hat{\pi}_n/\hat{\sigma}_{\pi,n})}{\hat{\sigma}_{\pi,n}\phi(\sqrt{n}\hat{\pi}_n/\hat{\sigma}_{\pi,n})}\biggl[\hat{\gamma}_n-\frac{\hat{\sigma}_{\gamma\pi,n}}{\hat{\sigma}_{\gamma,n}^2}\hat{\pi}_n\biggr]+\frac{\hat{\sigma}_{\gamma\pi,n}}{\hat{\sigma}_{\gamma,n}^2}
\]
where $\sigma$s are consistently estimated.
This is an example in which it makes sense to consider $T$ other than $\beta_{P,\psi}$ even if $\hat{\psi}_n\in\mathbb{D}_{P,\beta}$ almost surely.
\end{exa}

\begin{exa}[continues=exa:nonreg,name=Nonlinear regression]
The estimator $\hat{\beta}_n$ of $\beta$ that constitutes a minimizer $(\hat{\pi}_n,\hat{\beta}_n)$ of $\mathbb{E}_n[(y-cm(x;b))^2]$ with respect to $(c,b)$ is regular, provided that the estimator $\hat{\psi}_n\vcentcolon=\hat{\pi}_nm(\cdot;\hat{\beta}_n)$ of $\psi$ is regular.
\end{exa}

\begin{exa}[continues=exa:ngmm,name=Nonlinear GMM]
Let $\hat{m}_n(\cdot,\cdot)\vcentcolon=\mathbb{E}_n[M_i(\cdot,\cdot)]$ be a regular estimator of the moment function.
The GMM estimator for $(\pi,\beta)$ solves
\[
	\min_{c,b}\,\sqrt{n}\hat{m}_n(c,b)'W\sqrt{n}\hat{m}_n(c,b)
\]
for some $\ell\times\ell$ positive definite matrix $W$.
Oftentimes $W$ is estimated using initial estimates of $(\pi,\beta)$ or updated simultaneously with minimization.
In the former case, if $\hat{W}$ is continuous in the initial estimates, then the resulting GMM estimator is regular since minimization is continuous in the given norm.
In the latter case, if the whole objective function is continuous, the GMM estimator is regular.
%
\end{exa}

\subsection{Weakly Efficient Estimators}

We show that for any regular estimator of a weakly regular parameter, there exists another regular estimator that is weakly better in terms of convex loss.
A strict improvement is possible unless the estimator is already a nonrandom transformation of an efficient estimator of the underlying parameter.
%
For regular $\tilde{\beta}=S(\tilde{\psi})$, a particular improvement is given as the conditional expectation of $\tilde{\beta}$ conditional on an efficient estimator $\hat{\psi}$, that is, $\mathbb{E}[S(\tilde{\psi})\mid\hat{\psi}]$.
To formalize the efficiency gain, however, we use the normalized expression $\tilde{\beta}=T(\sqrt{n}(\tilde{\psi}-\psi(P)))$ for unknown $T$ and $\psi(P)$.

\begin{thm}[Local asymptotic Rao\hyp{}Blackwellization] \label{thm:lar}
Let $\beta:\mathcal{P}_\beta\to\mathbb{B}$ be weakly regular and $\psi:\mathcal{P}\to\mathbb{D}$ a minimal underlying regular parameter for $\beta$.
Let $\tilde{\psi}_n$ be a regular sequence of estimators of $\psi$ and $\tilde{\beta}_n=T(\sqrt{n}(\tilde{\psi}_n-\psi(P)))+o_P(1)$ be a regular sequence of estimators of $\beta$.
Suppose that an efficient regular sequence of estimators $\hat{\psi}_n$ of $\psi$ exists and $\bar{T}(\delta)\vcentcolon=\mathbb{E}[T(\delta+L_\eta/\sqrt{n})]$
exists as a Bochner integral.
Then $\hat{\beta}_n\vcentcolon=\bar{T}(\sqrt{n}(\hat{\psi}_n-\psi(P)))$ is a better regular estimator than $\tilde{\beta}_n$ in the sense that 
for every convex continuous loss function $\ell:\mathbb{B}\to\mathbb{R}$ such that $\ell(\tilde{\beta}_n-\beta(Q_n))$ and $\ell(\hat{\beta}_n-\beta(Q_n))$ are asymptotically equiintegrable under $Q_n\in\mathscr{P}_{P,\beta}$,%
\footnote{$X_n$ is {\em asymptotically equiintegrable} if $\lim_{M\to\infty}\limsup_{n\to\infty}\mathbb{E}^\ast[|X_n|\mathbbm{1}\{|X_n|>M\}]=0$ \Citep[p.\ 421]{vw1996}.}
\[
	\liminf_{n\to\infty\vphantom{P_P}} \, \mathbb{E}_{Q_n,\ast}[\ell(\tilde{\beta}_n-\beta(Q_n))]-\mathbb{E}_{Q_n}^\ast[\ell(\hat{\beta}_n-\beta(Q_n))] \geq 0.
\]
\end{thm}

Note that although $T$ and $\psi(P)$ are not known, we can construct a feasible estimator that is asymptotically equivalent to $\hat{\beta}$.
Let $\tilde{\beta}=S(\tilde{\psi})$ where $\tilde{\psi}$ is an inefficient estimator of $\psi$ that is asymptotically normal.
Then, the Rao\hyp{}Blackwellized estimator for $\beta$ is calculated as follows: (1) compute efficient $\hat{\psi}$ and $\Var(\tilde{\psi}\mid\hat{\psi})$; (2) compute $\hat{\beta}=\mathbb{E}_m[S(\hat{\psi}+e_j)\mid\hat{\psi}]$ where $e_1,\dots,e_m$ are drawn i.i.d.\ from $N(0,\Var(\tilde{\psi}\mid\hat{\psi}))$.
This feasible $\hat{\beta}$ has the same property as $\hat{\beta}$ in \cref{thm:lar}.



\begin{rem}
\cref{thm:lar} is a kind of admissibility requirement for a convex loss, e.g., it does not exclude constant estimators.
Unlike admissibility, however, it confines attention to the class of regular estimators while providing an improvement method of Rao\hyp{}Blackwellization.
If $\mathbb{B}=\mathbb{R}$, $\hat{\beta}_n$ first\hyp{}order stochastically dominates $\tilde{\beta}_n$.
\end{rem}


\begin{rem}
Efficiency is usually justified for {\em subconvex} loss functions (\Citealp[Theorem 3.11.5]{vw1996}). \cref{thm:lar} is in the same spirit but restricts us to {\em convex} functions.%
\footnote{Technically, there is no implication between convexity and subconvexity of a function. In this context, subconvexity can be thought of as roughly weaker.}
This difference comes from the fact that our best asymptotic distribution is a nonlinear transformation of Gaussian; there is no symmetry of the distribution we can exploit to accommodate subconvexity.
\end{rem}

Now we define our efficiency concept and introduce examples.

\begin{defn}[Weak efficiency for weakly regular parameter]
A regular sequence of estimators $\hat{\beta}_n$ for a weakly regular parameter $\beta$ is {\em weakly (semiparametric) efficient} at $P\in\mathcal{P}$ relative to $\mathscr{P}_{P,\beta}$ if the involved sequence of estimators for the minimal underlying regular parameter $\hat{\psi}_n$ is efficient.
\end{defn}


\begin{exa}[continues=exa:iv,name=Linear IV] \label{exa:iv3}
Suppose that the reduced\hyp{}form errors are heteroskedastic and the feasible GLS estimator is available.
Then, we can improve many estimators by \cref{thm:lar}, including even the optimal IV estimator.

Denote by $(\tilde{\gamma}_n,\tilde{\pi}_n)$ and $(\hat{\gamma}_n,\hat{\pi}_n)$ the OLS and GLS estimators of the reduced\hyp{}form coefficients.
By the efficiency of GLS,
\(
	\bigl[\begin{smallmatrix} \sqrt{n}(\tilde{\gamma}_n-\hat{\gamma}_n) \\ \sqrt{n}(\tilde{\pi}_n-\hat{\pi}_n) \end{smallmatrix}\bigr] \leadsto \begin{bsmallmatrix} e_\gamma \\ e_\pi \end{bsmallmatrix}
\).
Note that the asymptotic distributions of OLS and GLS are estimable, and GLS and the noise are independent, so we can consistently estimate the distribution of the noise.
Then, the Rao\hyp{}Blackwellized version of 2SLS takes the form $\bar{T}(\sqrt{n}\hat{\gamma}_n,\sqrt{n}\hat{\pi}_n)$ with
\[
	\bar{T}(\gamma,\pi)\vcentcolon=\mathbb{E}\bigl[\bigl([\pi+e_\pi]'\mathbb{E}[zz'][\pi+e_\pi]\bigr)^{-1}\bigl([\pi+e_\pi]'\mathbb{E}[zz'][\gamma+e_\gamma]\bigr)\bigr].
\]
This is weakly efficient since there is no more noise to Rao\hyp{}Blackwellize.
In practice, this expectation can be computed numerically.

To understand why the optimal IV can be improved, note that it exploits the heteroskedasticity of the structural error $\varepsilon$.
However, $\varepsilon$ cannot be consistently estimated while the reduced\hyp{}form errors $(u,v)$ can be.
This is where \cref{thm:lar} finds a room for improvement.
Since WLS is not as efficient as GLS, we can draw the noise and compute many instances of the optimal IV estimator,
\(
	\widehat{\bigl[\begin{smallmatrix}\pi\\&I_{dk}\end{smallmatrix}\bigr]}^{-1}\widehat{\bigl[\begin{smallmatrix}\gamma\\\vect(\pi)\end{smallmatrix}\bigr]},
\)
and take the numerical average to compute the Rao\hyp{}Blackwellized optimal IV estimator.


LIML is known to have no moment and CUE is suspected to have no moment, hence outside the scope of \cref{thm:lar}.
\end{exa}

\begin{exa}[continues=exa:nonreg,name=Nonlinear regression]
The form of the local parameter $\dot{\psi}_Pg=\mathbb{E}_P[\varepsilon g\mid x=\cdot]$ implies that $\varepsilon$ is an influence function, and the {\em efficient} influence function is given by $\tilde{\psi}_P=\mathbb{E}_P[\varepsilon^2\mid x=\cdot]^{-1}\varepsilon$.
Therefore, if there exists a consistent estimator for $\mathbb{E}[\varepsilon^2\mid x]$, then minimizing $\mathbb{E}_n[(y-cm(x;b))^2/\widehat{\mathbb{E}[\varepsilon^2\mid x]}]$ yields a more efficient estimator of $\psi$ than minimizing $\mathbb{E}_n[(y-cm(x;b))^2]$.%
\footnote{See \Citet[Example 25.66]{v1998}.}
If $x$ is discrete, then $\mathbb{E}_n[\hat{\varepsilon}^2\mid x]$ would yield a consistent estimator for $\mathbb{E}[\varepsilon^2\mid x]$ if $\hat{\pi}_n$ is consistent toward $0$; if $y$ is binary, then the functional form of $\mathbb{E}[\varepsilon^2\mid x]$ is fully determined by $\mathbb{E}[y\mid x]$, hence estimable; if $\mathbb{E}[\varepsilon^2\mid x]$ is smooth, we may use a series estimator as in \citet{n1994}.%
\footnote{Note that unlike \cref{exa:iv3} the structural error $\varepsilon$ can be consistently estimated since $cm$ can be.}
Given that, we can Rao\hyp{}Blackwellize the original estimator derived from minimizing $\mathbb{E}[(y-cm(x;b))^2]$.

Nonlinear least squares is used to estimate discrete choice models, for example, to avoid derivative calculation. Our method allows us to improve efficiency in such cases.
%
%
\end{exa}

\begin{exa}[continues=exa:ngmm,name=Nonlinear GMM]
The first part of our theory enables us to find out if there is any nuisance part in the moment function in each specific model (that is, if the moment function is minimal). Given that, it is often the case  that $\mathbb{E}_n[M(\cdot,\cdot)]$ is an efficient estimator of $\mathbb{E}[M(\cdot,\cdot)]$. Then, there is no noise left to Rao\hyp{}Blackwellize.
%
\end{exa}

Weak efficiency generalizes classical efficiency through a differentiable map to an almost everywhere continuous map.
It is therefore straightforward to construct estimators that are ``efficient'' under both strong and weak identification asymptotics.
If $\psi$ is sufficient and $T_n$ asymptotes to a continuous map under weak regularity and to an efficient differentiable map under regularity, the estimator $\hat{\beta}=T_n(\hat{\psi})$ is weakly efficient under weak regularity and efficient under regularity.
For example, the Rao\hyp{}Blackwellized 2SLS exhibits this property.
This is desirable since, often in practice, we do not know which asymptotics is a better approximation to the finite\hyp{}sample situation in hand.
Using such an estimator ensures maximal precision regardless of the ``correct'' asymptotics.

Finally, note that the point of weak efficiency is to exclude nuisance variation from an estimator, and the concept itself does not pin down a unique efficient distribution.
For example, constant estimators or any linear transformations of weakly efficient estimators are weakly efficient.
In some cases it is possible to impose additional restrictions to make a weakly efficient estimator unique.
In linear IV with $d=k=1$ where we know the sign of $\pi$, the Rao\hyp{}Blackwellized unbiased estimator is unique.
Since many nonlinear functions of Gaussian means admit unique unbiased estimators, it may be possible to uniquely pin down an unbiased weakly efficient estimator in many models.
However, unbiased estimators do not always exist (e.g., in linear IV when the sign of $\pi$ is not known).

\section{Simulation of Weak Efficiency in Linear IV Models} \label{sec:sim}

To illustrate weak efficiency, we conduct simulation of linear IV models (\cref{exa:iv}) with overidentified conditional moment restrictions and heteroskedasticity.
We consider discrete instruments so that we can estimate the heteroskedastic structure without imposing further assumptions.
This enables us to compute the optimal IV and the feasible reduced\hyp{}form GLS estimators.
We focus on two estimators, 2SLS and optimal IV, under weak and strong identification asymptotics.%

We let $d=1$ and $k=3$ so that 2SLS has a second moment.
The sample size is chosen to be $n=1{,}000$.
The instrument $z_i$ is uniformly distributed in $\{-1,1\}^3$, taking eight distinct combinations.
The errors $(\varepsilon_i,v_i)$ are drawn from a normal distribution with mean $0$ and variance depending on $z_i$ as \Cref{table:hetero};
this dependence is determined randomly at the beginning of the simulation.
The true parameters are given by $\beta=1$ and $\pi=(1,1,1)'/\sqrt{n}$ under weak identification (weakly regular $\beta$) and $\beta=1$ and $\pi=(1,1,1)'$ under strong identification (regular $\beta$).
Simulation runs for 5,000 iterations.
The heteroskedasticity\hyp{}adjusted concentration parameter $\mathbb{E}[\mathbb{E}[vv'\mid z]^{-1/2}\pi'zz'\pi\mathbb{E}[vv'\mid z]^{-1/2}]$ is $0.0075$ for weak identification and $7.5484$ for strong identification.

\begin{table}[t!]
\centering
\singlespacing
\caption{Heteroskedasticity of $(\varepsilon_i,u_i,v_i)$ given $z_i$. Since $u_i=\varepsilon_i+v_i'\beta$, the matrices are of rank 2.}
\label{table:hetero}
\small
\begin{tabular}{cc}
\hline\hline
$z_i'$ & $\Var((\varepsilon_i,u_i,v_i)\mid z_i)$ \\
\hline
\rule{0pt}{1.5em}%
$(-1,-1,-1)$ & $\begin{psmallmatrix}\phantom{-}7.32&\phantom{-}4.40&-2.91\\\phantom{-}4.40&\phantom{-}2.65&-1.75\\-2.91&-1.75&\phantom{-}1.16\end{psmallmatrix}$ \\
[.6em]
$(\phantom{-}1,-1,-1)$ & $\begin{psmallmatrix}\phantom{-}8.29&\phantom{-}3.74&-4.55\\\phantom{-}3.74&\phantom{-}\mathllap{1}3.41&\phantom{-}9.66\\-4.55&\phantom{-}9.66&\phantom{-}\mathllap{1}4.21\end{psmallmatrix}$ \\
[.6em]
$(-1,\phantom{-}1,-1)$ & $\begin{psmallmatrix}\phantom{-}3.78&\phantom{-}3.91&\phantom{-}0.14\\\phantom{-}3.91&\phantom{-}4.66&\phantom{-}0.74\\\phantom{-}0.14&\phantom{-}0.74&\phantom{-}0.61\end{psmallmatrix}$ \\
[.6em]
$(\phantom{-}1,\phantom{-}1,-1)$ & $\begin{psmallmatrix}\phantom{-}8.70&\phantom{-}3.60&-5.10\\\phantom{-}3.60&\phantom{-}4.54&\phantom{-}0.94\\-5.10&\phantom{-}0.94&\phantom{-}6.03\end{psmallmatrix}$ \\
[.5em]
\hline\hline
\end{tabular}
\quad
\begin{tabular}{cc}
\hline\hline
$z_i'$ & $\Var((\varepsilon_i,u_i,v_i)\mid z_i)$ \\
\hline
\rule{0pt}{1.5em}%
$(-1,-1,\phantom{-}1)$ & $\begin{psmallmatrix}\phantom{-}1.91&\phantom{-}1.77&-0.14\\\phantom{-}1.77&\phantom{-}2.19&\phantom{-}0.43\\-0.14&\phantom{-}0.43&\phantom{-}0.57\end{psmallmatrix}$ \\
[.6em]
$(\phantom{-}1,-1,\phantom{-}1)$ & $\begin{psmallmatrix}\phantom{-}3.83&-2.49&-6.32\\-2.49&\phantom{-}1.64&\phantom{-}4.14\\-6.32&\phantom{-}4.14&\phantom{-}\mathllap{1}0.46\end{psmallmatrix}$ \\
[.6em]
$(-1,\phantom{-}1,\phantom{-}1)$ & $\begin{psmallmatrix}\phantom{-}0.55&\phantom{-}0.32&-0.23\\\phantom{-}0.32&\phantom{-}0.20&-0.12\\-0.23&-0.12&\phantom{-}0.11\end{psmallmatrix}$ \\
[.6em]
$(\phantom{-}1,\phantom{-}1,\phantom{-}1)$ & $\begin{psmallmatrix}\phantom{-}1.22&\phantom{-}0.45&-0.77\\\phantom{-}0.45&\phantom{-}0.17&-0.28\\-0.77&-0.28&\phantom{-}0.49\end{psmallmatrix}$ \\
[.5em]
\hline\hline
\end{tabular}
\end{table}


To compute Rao\hyp{}Blackwellization, we must derive the feasible GLS estimator for the reduced\hyp{}form coefficients.
A nontrivial aspect of this is that it consists of multiple equations.
We handle this by combining them into one big equation:
\[
	\begin{bmatrix}Y\\\vect(X)\end{bmatrix}
	=\begin{bmatrix}Z&0\\0&\mathbbm{1}_d\otimes Z\end{bmatrix}\begin{bmatrix}\gamma\\\vect(\pi)\end{bmatrix}+\begin{bmatrix}U\\\vect(V)\end{bmatrix}
	=\vcentcolon\begin{bmatrix}Z&0\\0&\mathbbm{1}_d\otimes Z\end{bmatrix}\psi+\tilde{U}.
\]
Consequently, the variance\hyp{}covariance matrix of $\tilde{U}$ has some nonzero off\hyp{}diagonal elements.
We estimate it with initial OLS coefficients to compute the feasible GLS estimator for $(\gamma,\pi)$.
Since GLS is efficient, by orthogonality we have
$\Var(\hat{\psi}_{\text{OLS},n}-\hat{\psi}_{\text{GLS},n}\mid Z)=\Var(\hat{\psi}_{\text{OLS},n}\mid Z)-\Var(\hat{\psi}_{\text{GLS},n}\mid Z)$.
With this, we compute the conditional expectation of 2SLS conditional on GLS using $100$ draws from
$\begin{bsmallmatrix}e_{\gamma}\\e_\pi\end{bsmallmatrix} \sim N\bigl(\begin{bsmallmatrix}0\\0\end{bsmallmatrix},\,\Var(\hat{\psi}_{\text{OLS},n}-\hat{\psi}_{\text{GLS},n}\mid Z)\bigr)$.
In particular, the RB 2SLS estimator of $\beta$ is given by%
\footnote{Note that $e_\gamma$ and $e_\pi$ are already denormalized by $\sqrt{n}$.}
\[
	\hat{\mathbb{E}}_e[((\hat{\pi}_{\text{FGLS},n}+e_\pi)'(Z'Z)(\hat{\pi}_{\text{FGLS},n}+e_\pi))^{-1}(\hat{\pi}_{\text{FGLS},n}+e_\pi)'(Z'Z)(\hat{\gamma}_{\text{FGLS},n}+e_\gamma)],
\]
where $\hat{\mathbb{E}}_e$ denotes numerical expectation with respect to $(e_\gamma,e_\pi)$.

Rao\hyp{}Blackwellization of the optimal IV estimator requires a more elaborate procedure, as the optimal IV estimator involves two levels of noises.
The first noise comes from the fact that $\varepsilon$, needed to compute $A(z)$, cannot be consistently estimated; if it is estimated with the 2SLS residuals, then it contains noise due to inefficiency of OLS used in 2SLS.
The second noise comes from the fact that the optimal IV estimator is a function of the WLS estimator of $\bigl[\begin{smallmatrix}\pi\\&I_{dk}\end{smallmatrix}\bigr]$ and $\bigl[\begin{smallmatrix}\gamma\\\vect(\pi)\end{smallmatrix}\bigr]$, where the weights are given by the estimated $A(z)$.
We use 50 draws to Rao\hyp{}Blackwellize the first noise and 100 draws for each of the first noise to Rao\hyp{}Blackwellize the second.


\begin{figure}[t!]
\centering
\begin{subfigure}[t]{0.45\textwidth}
\centering
\includegraphics[width=\textwidth]{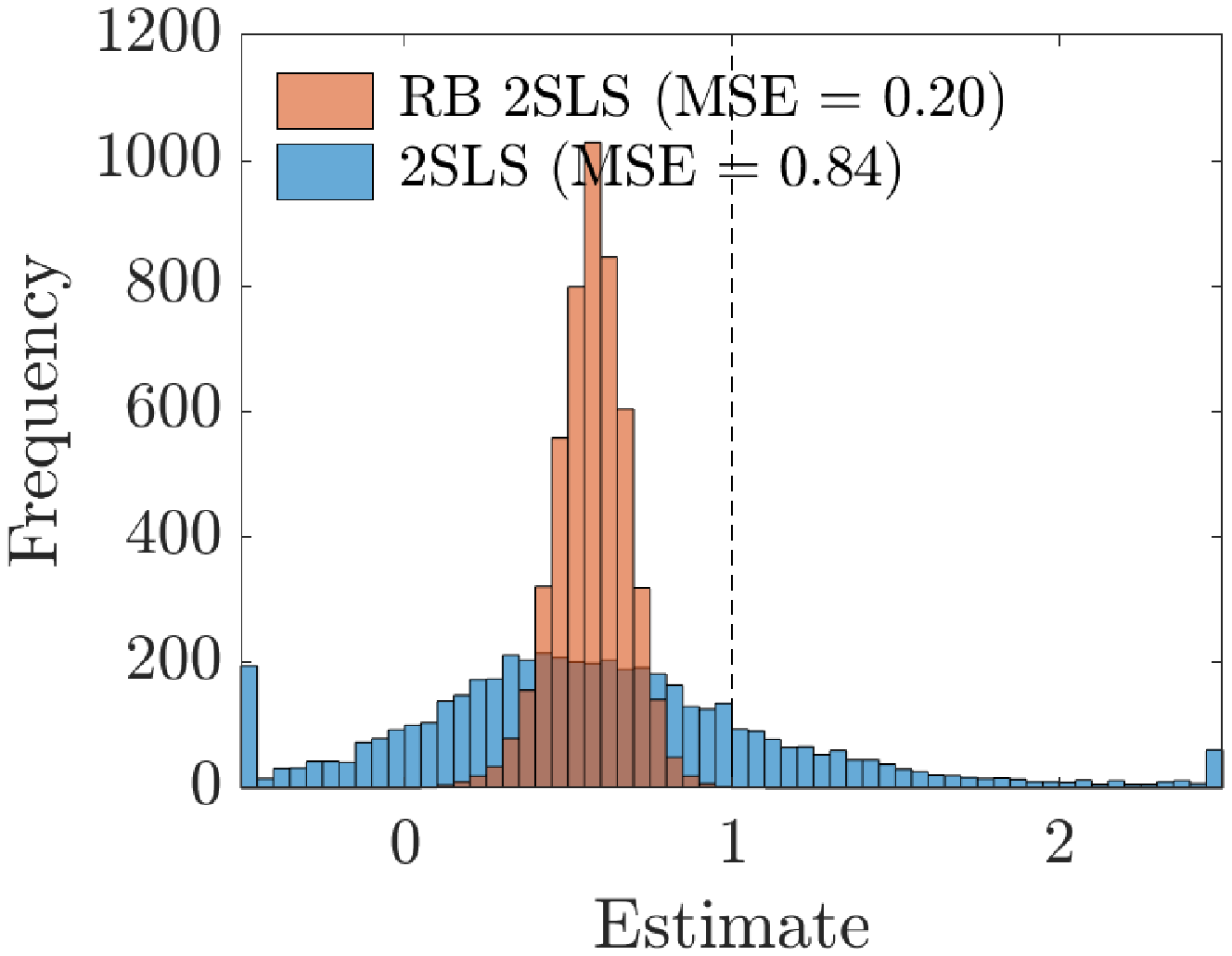}
\caption{Histograms of 2SLS and RB 2SLS estimators.}
\label{fig:1a}
\end{subfigure}
\qquad
\begin{subfigure}[t]{0.45\textwidth}
\centering
\includegraphics[width=\textwidth]{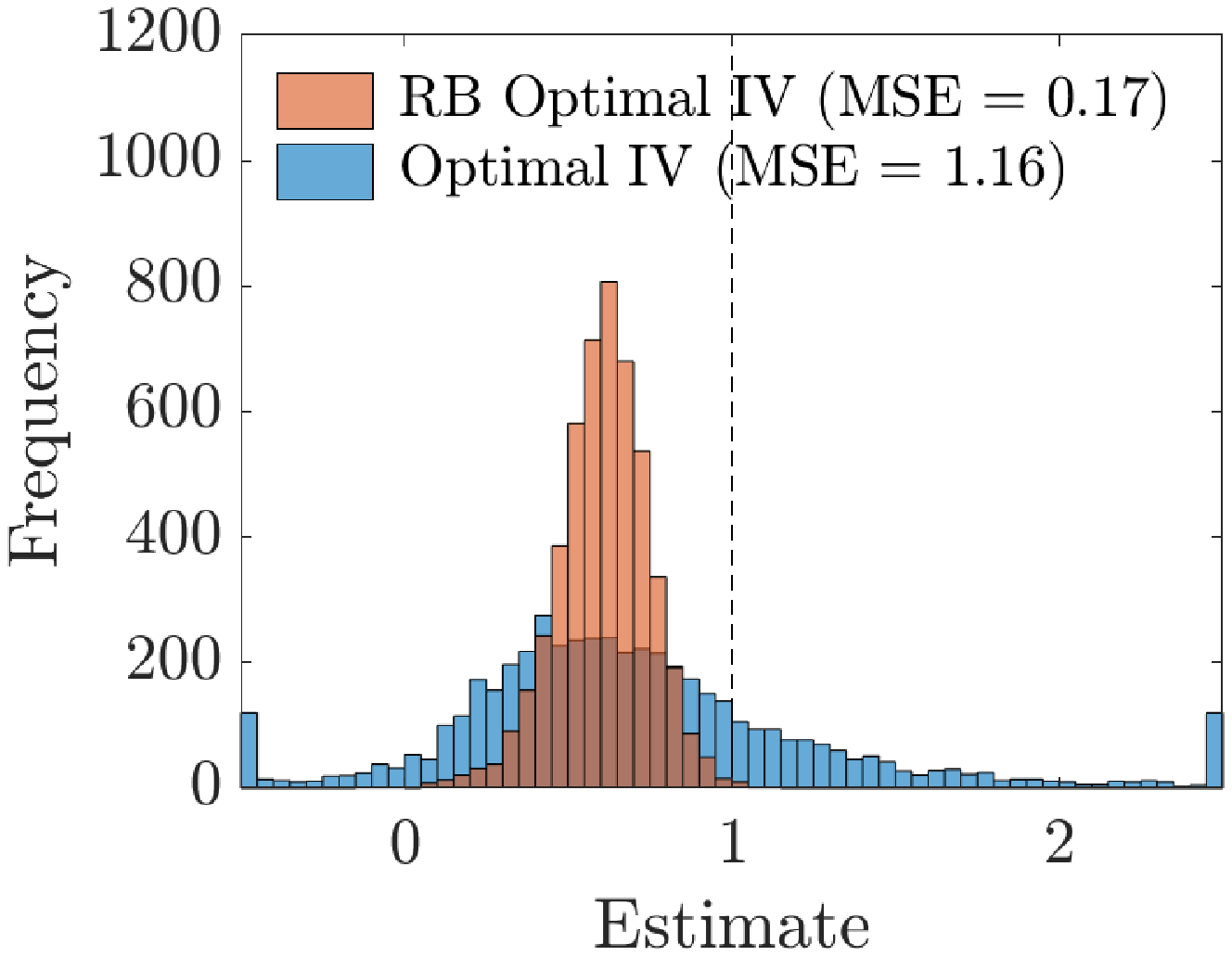}
\caption{Histograms of optimal IV and RB optimal IV estimators.}
\label{fig:1b}
\end{subfigure}
\caption{Distributions of 2SLS, optimal IV, and their Rao\hyp{}Blackwellization {\em under weak regularity of $\beta$ (weak identification asymptotics)}. Rao\hyp{}Blackwellization improves the mean squared errors. Simulated with 1,000 observations and 5,000 iterations. Clusters at the boundaries indicate observations outside the range.}
\label{fig:1}
\end{figure}

\Cref{fig:1a} is the histograms of 2SLS and RB 2SLS estimators under weak regularity of $\beta$.
The vertical dotted line indicates the true value, $\beta=1$.
It shows that the distribution of RB 2SLS is more concentrated than 2SLS.
Since Rao\hyp{}Blackwellization does not affect its mean, both estimators share the same bias.
\Cref{fig:1b} is the histograms of optimal IV and RB optimal IV estimators for the same run, in which we observe similar results.
To connect these histograms to \cref{thm:lar}, we consider two loss functions: the mean squared error (MSE) $\ell(x)=x^2$ and the mean absolute error (MAE) $\ell(x)=|x|$, as summarized in \Cref{table1}.
The MSE of 2SLS decreases from $0.84$ to $0.20$ after Rao\hyp{}Blackwellization; the MSE of optimal IV from $1.16$ to $0.17$.
The MAE of 2SLS and optimal IV shows similar drop.
We see substantial decrease in the losses in both estimators.
LAR (\cref{thm:lar}) guarantees that the losses of the RB versions do not exceed those of the original ones, at least asymptotically.
In this sense, it is preferable to use a weakly efficient estimator whenever available.

\begin{figure}[t!]
\centering
\begin{subfigure}[t]{0.45\textwidth}
\centering
\includegraphics[width=\textwidth]{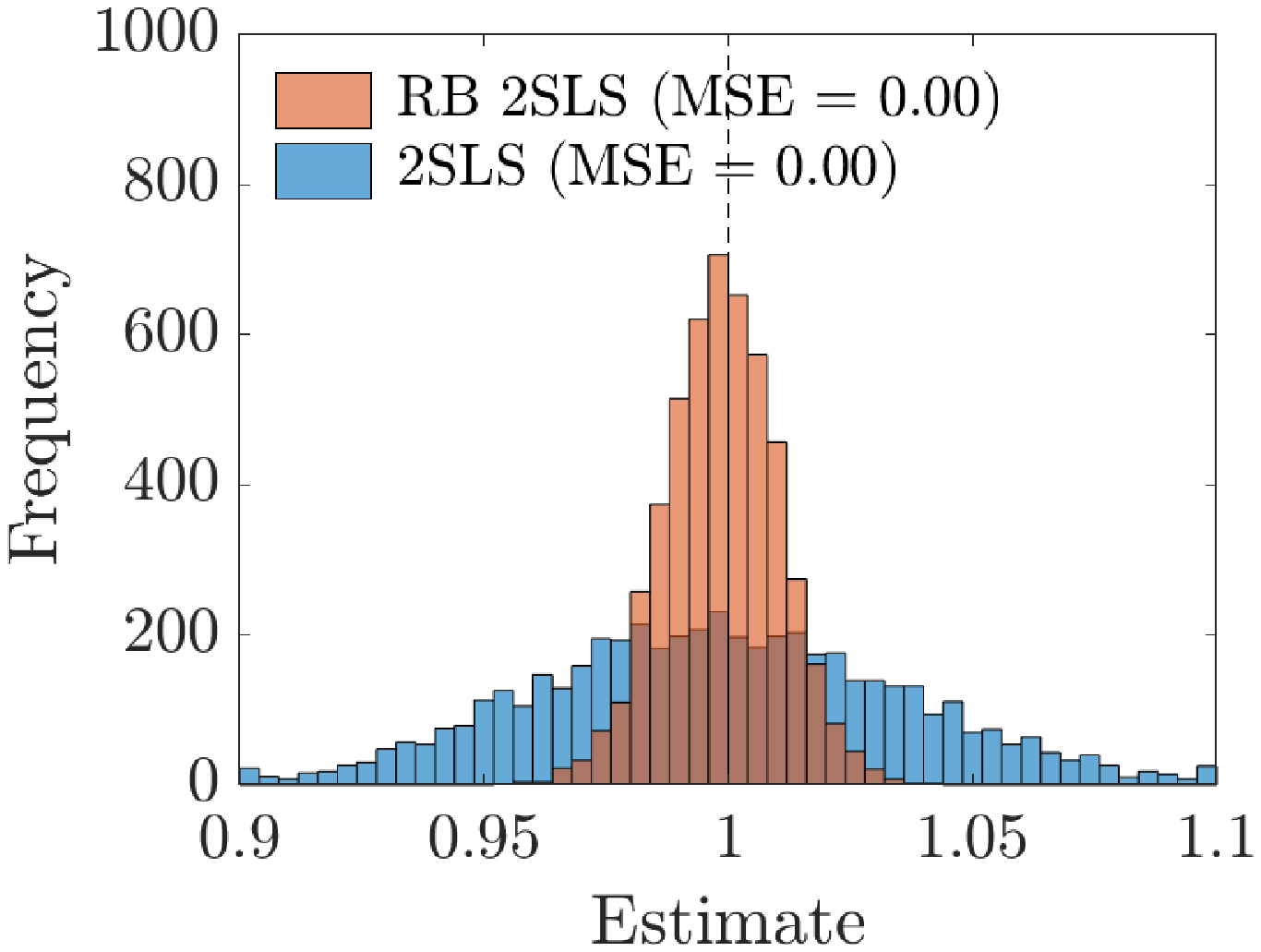}
\caption{Histograms of 2SLS and RB 2SLS estimators.}
\label{fig:2a}
\end{subfigure}
\qquad
\begin{subfigure}[t]{0.45\textwidth}
\centering
\includegraphics[width=\textwidth]{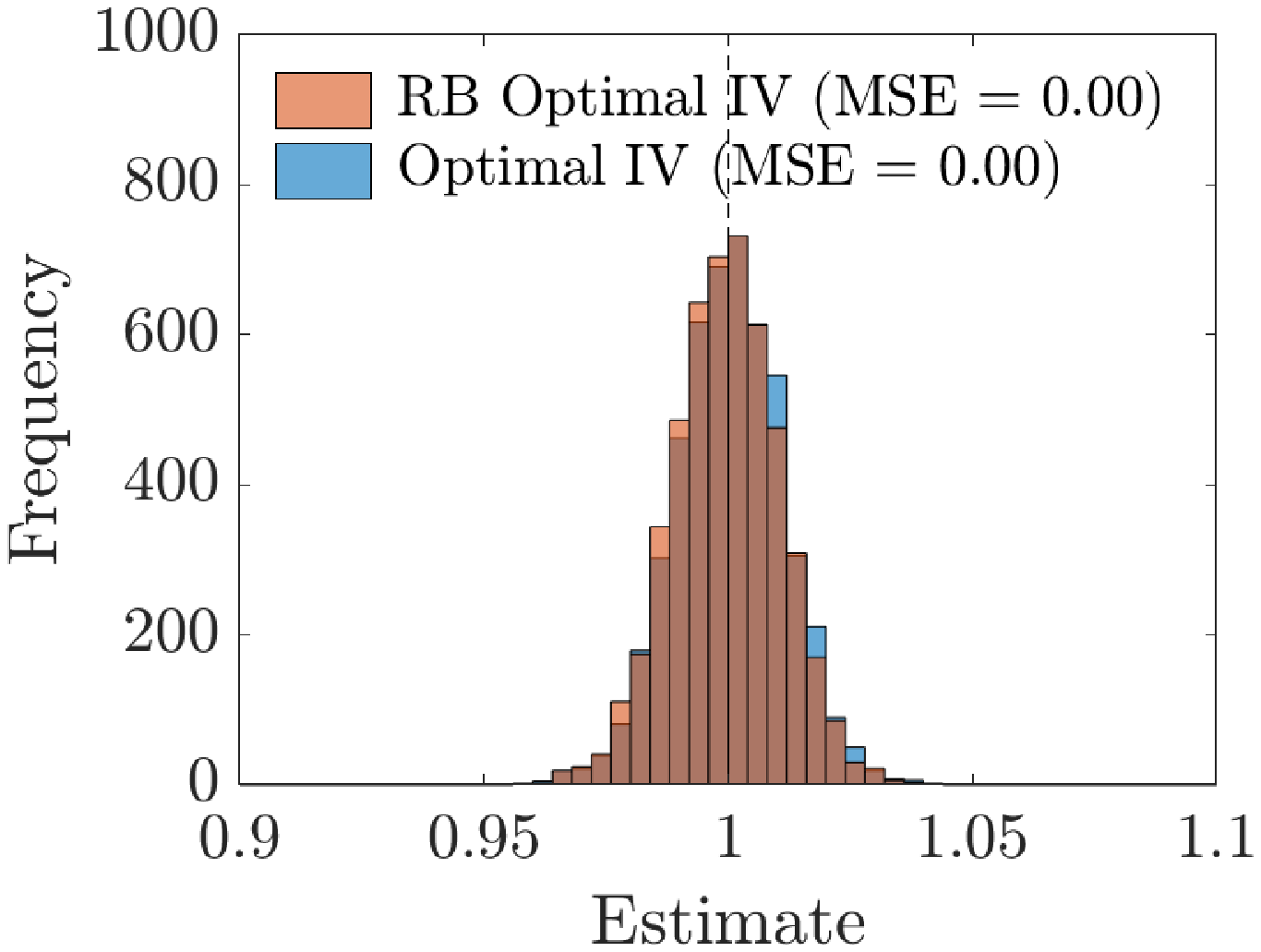}
\caption{Histograms of optimal IV and RB optimal IV estimators.}
\label{fig:2b}
\end{subfigure}
\caption{Distributions of 2SLS, optimal IV, and their Rao\hyp{}Blackwellization {\em under regularity of $\beta$ (strong identification asymptotics)}. Asymptotic distribution of RB 2SLS coincides with the optimal IV. Simulated with 1,000 observations and 5,000 iterations. Clusters at the boundaries indicate observations outside the range.}
\label{fig:2}
\end{figure}

\Cref{fig:2} is the histograms of the same estimators under regularity of $\beta$.
From classical results, we know that optimal IV is efficient and 2SLS is not.
We see that both RB optimal IV and RB 2SLS coincide with optimal IV under strong identification asymptotics.
This suggests that LAR does not alter an already efficient estimator while it transforms an inefficient estimator into an efficient one.
The condition for this to hold can be understood using the analogy introduced at the beginning of \cref{sec:lax}.
For an estimator of the form $\hat{\beta}=T_n(\hat{\psi})$, if $T_n$ asymptotes to an almost everywhere continuous map under weak regularity of $\beta$ and to the optimal $T$ under regularity of $\beta$, then $\hat{\beta}$ is weakly efficient under weak identification asymptotics and efficient under strong identification asymptotics.
This applies to most of known regular estimators.

Computational time of Rao\hyp{}Blackwellization does not necessarily parallel computational time of the original estimator.
There are two components that contribute to computational burden, $T_n$ and $\hat{\psi}_n$, and Rao\hyp{}Blackwellization only repeats $T_n$.
In our simulation, therefore, Rao\hyp{}Blackwellization is done very quick.
In fact, the most time\hyp{}consuming part of our simulation is the computation of the original optimal IV estimator, for which derivation of $\hat{\psi}_n$ requires a loop of matrix operations over observations.
In our laptop, one iteration of the simulation (computation of 2SLS, optimal IV, their Rao\hyp{}Blackwellization, and some auxiliary computation) takes less than 0.3 seconds.
From a standpoint of strong identification asymptotics, our RB estimators (or the 2SLS with GLS estimators in place of OLS) give a much faster way to compute efficient estimators than to compute optimal IV.

\begin{table}[t!]
\centering
\singlespacing
\caption{The MSE and MAE of 2SLS, optimal IV, and their Rao\hyp{}Blackwellizations under weak and strong identification asymptotics.}
\label{table1}
\small
\begin{tabular}{lccccccccccc}
\hline\hline
& \multicolumn{5}{c}{\em Weak Identification} && \multicolumn{5}{c}{\em Strong Identification} \\
\cline{2-6} \cline{8-12}
& \multicolumn{2}{c}{2SLS} && \multicolumn{2}{c}{Optimal IV} && \multicolumn{2}{c}{2SLS} && \multicolumn{2}{c}{Optimal IV} \\
\cline{2-3} \cline{5-6} \cline{8-9} \cline{11-12}
& Plain & RB && Plain & RB && Plain & RB && Plain & RB \\
\hline
\rule{0pt}{1.2em}MSE & 0.841 & 0.196 && 1.159 & 0.174 && 0.001 & 0.000 && 0.000 & 0.000 \\
MAE & 0.633 & 0.428&& 0.604 & 0.394 && 0.030 & 0.009 && 0.009 & 0.009 \\
[.2em]
\hline
Observations & \multicolumn{2}{c}{1,000} && \multicolumn{2}{c}{1,000} && \multicolumn{2}{c}{1,000} && \multicolumn{2}{c}{1,000} \\
Noise draws & \multicolumn{2}{c}{100} && \multicolumn{2}{c}{$50\times100$} && \multicolumn{2}{c}{100} && \multicolumn{2}{c}{$50\times100$} \\
Iterations & \multicolumn{2}{c}{5,000} && \multicolumn{2}{c}{5,000} && \multicolumn{2}{c}{5,000} && \multicolumn{2}{c}{5,000} \\
\hline\hline
\end{tabular}
\end{table}

Note that the conditional moment restrictions, $\mathbb{E}[u_i\mid z_i]=0$ and $\mathbb{E}[v_i\mid z_i]=0$, play a crucial role in this exercise.
OLS is inefficient because of them.
Relating thereto, another important assumption is the availability of feasible GLS.
A notable example in which the form of heteroskedasticity is known {\em a priori} is when $y_i$ is binary and one has a conditional moment restriction, $\mathbb{E}[y_i\mid x_i]=f(x_i)$; the form of heteroskedasticity is uniquely determined by $f$ as $\mathbb{E}[(y_i-f(x_i))^2\mid x_i]=f(x_i)-f(x_i)^2$. 
If $f$ can be estimated, for example when $f$ belongs to some parametric family $\{f_\theta\}$, one may use feasible GLS with no additional loss of generality.
In other linear models with an unknown form of heteroskedasticity, feasible GLS with a nonparametric estimator is available under various assumptions \citep{r1987,n1994}.


%
%


\section{Conclusion} \label{sec:conclusion}

This paper studies weak identification in semiparametric models and investigates efficient estimation. 
Weak identification is captured by the notion of {\em weak regularity}, with which the parameter is approximated by a homogeneous\hyp{}of\hyp{}degree\hyp{}zero map of the score.
This nonlinearity implies impossibility of consistent estimation and inference and equivariant estimation.
For each weakly regular parameter, there exists an underlying parameter that is regular and fully characterizes the weakly regular parameter locally.
An underlying parameter that is minimal and sufficient shares the same nuisance tangent space with the weakly regular parameter, representing the exact amount of information relevant to the weakly regular parameter.

Decomposing estimation of the weakly regular parameter into estimation of the minimal underlying parameter and its transformation, efficiency is discussed in terms of the noise involved in the estimator of the underlying regular parameter.
When the estimator of the underlying parameter is inefficient, we can improve the estimator of the weakly regular parameter by taking expectation conditional on the efficient estimator of the underlying parameter.
The estimator is called {\em weakly efficient} if no further improvement is possible.
This exploits the property that an efficient estimator of a regular parameter is ``asymptotically sufficient", hence the name {\em local asymptotic Rao\hyp{}Blackwellization}.
Simulation of the linear IV model demonstrates that the 2SLS and optimal IV estimators can be improved if the feasible GLS estimator of the reduced\hyp{}form coefficients is available.




%
%


\section*{Appendix}

\appendix
\section{Proofs} \label{sec:proofs}


\begin{proof}[Proof of \cref{lem:cone}]
Since $\dot{\mathcal{P}}_P$ is assumed to be linear, if $g\in\dot{\mathcal{P}}_P$ then $ag\in\dot{\mathcal{P}}_P$ for every $a\in\mathbb{R}$. If $g$ is induced by a path $t\mapsto Q_t$ and $a>0$, then $ag$ can be induced by the path $t\mapsto Q_{at}$, which is the same path up to a scaled index. Therefore, if $Q_t\in\mathscr{P}_P\setminus\mathscr{P}_{P,\beta}$ then $Q_{at}\in\mathscr{P}_P\setminus\mathscr{P}_{P,\beta}$, implying that if $g\in\dot{\mathcal{P}}_P\setminus\dot{\mathcal{P}}_{P,\beta}$ then $ag\in\dot{\mathcal{P}}_P\setminus\dot{\mathcal{P}}_{P,\beta}$. Being defined as a difference between a linear space and a cone, $\dot{\mathcal{P}}_{P,\beta}$ is a cone.
\end{proof}

\begin{proof}[Proof of \cref{thm:impossibility}]
Let $\beta:\mathcal{P}_\beta\to\mathbb{B}$ be weakly regular and $\beta_P$ nonconstant.

{\em The first assertion.}
Suppose that $\hat{\beta}_n:\mathcal{X}^n\to\mathbb{B}$ is a consistent sequence of estimators, or even weaker, that there exist two paths $Q_{n1},Q_{n2}\in\mathscr{P}_{P,\beta}$ inducing $g_1,g_2\in\dot{\mathcal{P}}_{P,\beta}$ such that $\beta_P(g_1)\neq\beta_P(g_2)$ and $\hat{\beta}_n\conv^{Q_{nj}\ast}\beta_P(g_j)$ under each $Q_{nj}\in\{Q_{n1},Q_{n2}\}$.
Define $2\varepsilon\vcentcolon=\|\beta_P(g_1)-\beta_P(g_2)\|_{\mathbb{B}}$.
Denote by $Q_{nj}^n$ the product measure of $Q_{nj}$ on the product sample space $\mathcal{X}^n$.
By the portmanteau theorem \Citep[Theorem 1.3.4]{vw1996} and the assumption of convergence in outer probability,
$\limsup_{n\to\infty}Q_{n1}^n(\|\hat{\beta}_n-\beta_P(g_1)\|_{\mathbb{B}}^\ast\geq\varepsilon)\leq0$
while
$\liminf_{n\to\infty}Q_{n2}^n(\|\hat{\beta}_n-\beta_P(g_1)\|_{\mathbb{B}}^\ast\geq\varepsilon)\geq\liminf_{n\to\infty}Q_{n2}^n(\|\hat{\beta}_n-\beta_P(g_1)\|_{\mathbb{B},\ast}>\varepsilon)\geq1$.
Therefore, $Q_{n2}^n$ is not contiguous to $Q_{n1}^n$. 
Being paths, however, $Q_{n2}^n$ must be contiguous to $P^n$ and $P^n$ to $Q_{n1}^n$ \Citep[Lemma 3.10.11 and Theorem 3.10.9]{vw1996}, hence a contradiction.

{\em The second assertion.}
Let $H_0:\beta\in\mathbb{B}_0$ and $H_1:\beta\in\mathbb{B}_1$ be the null and alternative hypotheses such that $\mathbb{B}_0$ and $\mathbb{B}_1$ are nonempty.
Suppose that $\phi_n:\mathcal{X}^n\to[0,1]$ is a consistent sequence of tests of $H_0$ of level $\alpha<1$ so that there exist two paths $Q_{n0},Q_{n1}\in\mathscr{P}_{P,\beta}$ with $\beta_P(g_0)\in\mathbb{B}_0$ and $\beta_P(g_1)\in\mathbb{B}_1$ such that $\phi_n\to^{Q_{n0}\ast}\alpha$ and $\phi_n\to^{Q_{n1}\ast}1$.
Then by the same reasoning a contradiction follows.

{\em The third assertion.}
Let $\hat{\beta}_n$ be an equivariant\hyp{}in\hyp{}law sequence of estimators of $\beta$ with a separable limit law, that is, there exists a fixed separable Borel probability measure $L$ on $\mathbb{B}$ such that
$\hat{\beta}_n-\beta(Q_n)\overset{Q_n}{\leadsto}L$ for every $Q_n\in\mathscr{P}_{P,\beta}$.
We derive contradiction by constructing two paths along which $\beta$ takes distinct values but the likelihood ratio of which converges to $1$; this means that $\hat{\beta}$ follows the same distribution in both paths by Le Cam's third lemma; therefore, $\hat{\beta}-\beta$ must follow different distributions.
Pick $g_1,g_2\in\dot{\mathcal{P}}_{P,\beta}$ such that $\beta_P(g_1)\neq\beta_P(g_2)$ and denote $\beta_1\vcentcolon=\beta_P(g_1)$ and $\beta_2\vcentcolon=\beta_P(g_2)$.
Since $\dot{\mathcal{P}}_{P,\beta}$ is a cone (\cref{lem:cone}), $ag_1$ and $ag_2$ are also in $\dot{\mathcal{P}}_{P,\beta}$ for every $a>0$ and by homogeneity we have $\beta_P(ag_j)=\beta_j$.
For each positive integer $k$, take $Q_{nk1},Q_{nk2}\in\mathscr{P}_{P,\beta}$ to be paths that induce scores $g_1/k$ and $g_2/k$.
Let $d_{Q_n}$ denote the metric that metrizes weak topology on $\mathbb{B}$ under $Q_n$ toward separable limits \Citep[p.\ 73]{vw1996}.
For each $k$, let $n_k$ be such that for every $n\geq n_k$,
\begin{gather*}
	\int_{\mathcal{X}} \biggl[ \frac{dQ_{nk1}^{1/2}-dP^{1/2}}{1/\sqrt{n}}-\frac{1}{2}\frac{g_1}{k}dP^{1/2} \biggr]^2 \vee \int_{\mathcal{X}} \biggl[ \frac{dQ_{nk2}^{1/2}-dP^{1/2}}{1/\sqrt{n}}-\frac{1}{2}\frac{g_2}{k}dP^{1/2} \biggr]^2 < \frac{1}{k}, \\
	d_{Q_{nk1}}\bigl(\hat{\beta}_n-\beta(Q_{nk1}),L\bigr)\vee d_{Q_{nk2}}\bigl(\hat{\beta}_n-\beta(Q_{nk2}),L\bigr)<\frac{1}{k}.
\end{gather*}
Then one can take $n_k'$ so that $n_k'\geq n_k$ and $n_{k+1}'>n_k'$ for every $k$.
Construct two paths $Q_{n1}'$ and $Q_{n2}'$ by $Q_{nj}'=Q_{nk_nj}$ where $k_n$ satisfies $n_{k_n}'\leq n<n_{k_n+1}'$.
Then $Q_{nj}'\to^{\text{DQM}}P$ with scores equal to zero and $\hat{\beta}_n-\beta(Q_{nj}')$ converges weakly to $L$ under $Q_{nj}'$.
Now we want to show that $dQ_{n2}'^n/dQ_{n1}'^n$ converges to $1$ and invoke Le Cam's third lemma.
For this, we adopt the same proof strategy as \Citet[Theorem 7.2]{v1998}.
Observe that
$\mathbb{E}_{Q_{n1}'} \bigl[ n \bigl( 1-\frac{dQ_{n2}'^{1/2}}{dQ_{n1}'^{1/2}} \bigr)^2 \bigr]\leq\int_{\mathcal{X}} \bigl[ \frac{dQ_{n1}'^{1/2}-dQ_{n2}'^{1/2}}{1/\sqrt{n}} \bigr]^2 \conv 0$.
By Taylor's theorem, $\log x^2=-2(1-x)-(1-x)^2+(1-x)^2R(1-x)$ for $R:\mathbb{R}\to\mathbb{R}$ such that $R(1-x)\to0$ as $x\to1$. Then,
\(
	\log\frac{dQ_{n2}'^n}{dQ_{n1}'^n}(X_1,\dots,X_n)
	=\log\bigl(\frac{dQ_{n2}'}{dQ_{n1}'}(X_1)\cdots\frac{dQ_{n2}'}{dQ_{n1}'}(X_n)\bigr)
	=\sum_{i=1}^n\log\frac{dQ_{n2}'}{dQ_{n1}'}
	=-2\sum_{i=1}^n W_{ni}-\sum_{i=1}^n W_{ni}^2+\sum_{i=1}^n W_{ni}^2 R(W_{ni})
\)
where $W_{ni}\vcentcolon=1-dQ_{n2}'^{1/2}/dQ_{n1}'^{1/2}(X_i)$.
We argue that all three terms converge to zero in probability.
Under $Q_{n1}'$,
\begin{gather*}
	\Biggl|\mathbb{E}\sum_{i=1}^n W_{ni}\Biggr|=n\Biggl|1-\int\frac{dQ_{n2}'^{1/2}}{dQ_{n1}'^{1/2}}dQ_{n1}'\Biggr|\leq \frac{1}{2}\int\biggl[\frac{dQ_{n1}'^{1/2}-dQ_{n2}'^{1/2}}{1/\sqrt{n}}\biggr]^2\conv0, \\
	\Var\Biggl(\sum_{i=1}^n W_{ni}\Biggr)\leq\mathbb{E}[nW_{ni}^2]=\mathbb{E}\biggl[n\biggl(1-\frac{dQ_{n2}'^{1/2}}{dQ_{n1}'^{1/2}}\biggr)^2\biggr]\conv0.
\end{gather*}
These results imply that the expectation and variance of $\sum W_{ni}$ converge to zero; hence it converges to zero in probability.
The second result implies that $nW_{ni}^2$ converges to zero in mean; by the law of large numbers $\sum W_{ni}^2$ converges to zero in probability.
By Markov's inequality,
\(
	\Pr(\max_{1\leq i\leq n}|W_{ni}|>\varepsilon)\leq n\Pr(|W_{ni}|>\varepsilon)\leq n\Pr(nW_{ni}^2>n\varepsilon^2)\leq\frac{\mathbb{E}[nW_{ni}^2]}{\varepsilon^2}\conv0
\)
for every $\varepsilon>0$.
Thus, $\max_{1\leq i\leq n}|W_{ni}|$ converges to zero in probability, and so does $\max_{1\leq i\leq n}|R(W_{ni})|$.
Therefore, the third term $\sum W_{ni}^2 R(W_{ni})$ converges to zero in probability.
We conclude that $dQ_{n2}'^n/dQ_{n1}'^n$ converges to $1$ in probability under $Q_{n1}'$.
Since $L$ is separable, by Slutsky's lemma \Citep[Example 1.4.7]{vw1996},
\(
	\bigl( \hat{\beta}_n, \frac{dQ_{n2}'^n}{dQ_{n1}'^n} \bigr) \overset{Q_{n1}'}{\leadsto} (\beta_1+L,1)
\).
By Le Cam's third lemma \Citep[Theorem 3.10.7]{vw1996},
\(
	(\beta_2+L)(B)=\mathbb{E}\mathbbm{1}\{\beta_1+L\in B\}1=(\beta_1+L)(B)
\)
for every Borel $B\subset\mathbb{B}$, which contradicts $\beta_1\neq\beta_2$.
%
\end{proof}

\begin{proof}[Proof of \cref{thm:exist}]
Denote by $\mathbb{D}$ the Banach space of $P$\hyp{}square integrable functions on $\mathcal{X}$ and define $\psi:\mathcal{P}\to\mathbb{D}$ by $\psi(Q)=dQ^{1/2}/dP^{1/2}$.
Note that $\psi$ is regular with derivative $\dot{\psi}_P:\dot{\mathcal{P}}_P\to\mathbb{D}$,
$\dot{\psi}_Pg=g$.
Thus, we have $\beta_{P,\psi}=\beta_P$.
\end{proof}

\begin{proof}[Proof of \cref{lem:sil}]
(i) Trivially, $0\in N(\beta_P)$. By definition, $\tilde{g}_1,\tilde{g}_2\in N(\beta_P)$ implies $\tilde{g}_1+\tilde{g}_2\in N(\beta_P)$. Take $\tilde{g}\in N(\beta_P)$ and $a>0$. Since $\dot{\mathcal{P}}_{P,\beta}$ is a cone (\cref{lem:cone}) and $\beta_P$ is homogeneous of degree zero, $\beta_P(g)=\beta_P(g/a)=\beta_P(g/a+\tilde{g})=\beta_P(g+a\tilde{g})$ for every $g\in\dot{\mathcal{P}}_{P,\beta}$. This means $a\tilde{g}\in N(\beta_P)$. Therefore, $N(\beta_P)$ is linear.
%
(ii) If $P\in\mathcal{P}\setminus\mathcal{P}_\beta$, then $0\notin\dot{\mathcal{P}}_{P,\beta}$. Since $g\in N(\beta_P)\cap\dot{\mathcal{P}}_{P,\beta}$ implies $\beta_P(g)=\beta_P(g-g)=\beta_P(0)$, $N(\beta_P)\cap\dot{\mathcal{P}}_{P,\beta}$ must be empty.
%
%
(iii) 
If $\Pi_\beta g=0$ then $g\in N(\beta_P)$, which implies $g\notin\dot{\mathcal{P}}_{P,\beta}$ by (ii).
\end{proof}

\begin{proof}[Proof of \cref{thm:minexist}]
Let $\mathbb{D}=L_2(P)$ and define $\psi:\mathcal{P}\to\mathbb{D}$ by $\psi(Q)=2\Pi_\beta dQ^{1/2}/dP^{1/2}$. Then $\psi$ is regular with the derivative $\dot{\psi}_P:\dot{\mathcal{P}}_P\to\mathbb{D}$,
$\dot{\psi}_Pg=\Pi_\beta g$.
Note that $\beta_P(g)=\beta_P(\Pi_\beta g)$.
This implies that $\psi$ is an underlying regular parameter for $\beta$ and that $N(\dot{\psi}_P)=N(\beta_P)$, which implies minimal sufficiency of $\psi$.
\end{proof}

\begin{proof}[Proof of \cref{thm:minchara}]
{\em Sufficiency.} Assume that for any sufficient underlying regular parameter $\phi:\mathcal{P}\to\mathbb{E}$ for $\beta$ there exists a map $\tau:\mathbb{E}\to\mathbb{D}$ such that $\tau(\dot{\phi}_Pg)=\dot{\psi}_Pg$ for every $g\in\dot{\mathcal{P}}_{P,\beta}$.
This means that $N(\dot{\phi}_P)\subset N(\dot{\psi}_P)$. Take $\phi$ to be minimal; then $N(\beta_P)=N(\dot{\phi}_P)\subset N(\dot{\psi}_P)$. On the other hand, since $\psi$ is assumed to be a sufficient underlying parameter, we have $N(\beta_P)\supset N(\dot{\psi}_P)$.

{\em Necessity.} Assume that $\psi:\mathcal{P}\to\mathbb{D}$ is a minimal sufficient underlying regular parameter for $\beta$. Take $\phi:\mathcal{P}\to\mathbb{E}$ to be another sufficient underlying regular parameter for $\beta$.
Then $\beta_{P,\psi}(\dot{\psi}_Pg)=\beta_{P,\phi}(\dot{\phi}_Pg)$ for every $g\in\dot{\mathcal{P}}_{P,\beta}$ and $N(\dot{\psi}_P)=N(\beta_P)\supset N(\dot{\phi}_P)$.
The first property implies $\dot{\psi}_Pg\in\beta_{P,\psi}^{-1}\beta_{P,\phi}(\dot{\phi}_Pg)$ for every $g\in\dot{\mathcal{P}}_{P,\beta}$.
The second property implies that if $\dot{\phi}_Pg_1=\dot{\phi}_Pg_2$ then $\dot{\psi}_Pg_1=\dot{\psi}_Pg_2$.
Conclude that there exists a map $\tau:\mathbb{E}_0\to\mathbb{D}$ such that $\dot{\psi}_Pg=\tau(\dot{\phi}_Pg)$ for $g\in\dot{\mathcal{P}}_0$ where $\mathbb{E}_0\vcentcolon=\dot{\phi}_P(\dot{\mathcal{P}}_{P,\beta})$.
Since $\dot{\phi}_P$ and $\dot{\psi}_P$ are linear in $g$, $\tau$ must be linear.
Finally, one can extend $\tau$ on the whole of $\mathbb{E}$ by letting $\tau(e)\vcentcolon=\tau(\Pi_{\mathbb{E}_0}e)$.
\end{proof}

\begin{proof}[Proof of \cref{thm:dist}]
The claim follows by the extended continuous mapping theorem \Citep[Theorem 1.11.1 and Problem 1.11.1]{vw1996}.
\end{proof}

\begin{proof}[Proof of \cref{thm:lar}]
Since expectation carries over continuity, $\hat{\beta}_n$ is regular.
Write
\[
\begin{multlined}
	\mathbb{E}_\ast[\ell(\tilde{\beta}_n-\beta)]-\mathbb{E}^\ast[\ell(\hat{\beta}_n-\beta)]
	=\mathbb{E}_\ast[\ell(\tilde{\beta}_n-\beta)]-\mathbb{E}[\ell(T(\dot{\psi}_Pg+L_\psi+L_\eta)-\beta)] \\
\begin{aligned}
	&\hphantom{={}}+\mathbb{E}[\mathbb{E}[\ell(T(\dot{\psi}_Pg+L_\psi+L_\eta)-\beta)-\ell(\bar{T}(\dot{\psi}_Pg+L_\psi)-\beta)\mid L_\psi]] \\
	&\hphantom{={}}+\mathbb{E}[\ell(\bar{T}(\dot{\psi}_Pg+L_\psi)-\beta)]-\mathbb{E}^\ast[\ell(\hat{\beta}_n-\beta)].
\end{aligned}
\end{multlined}
\]
The first and third differences converge to zero by \cref{thm:dist} and \Citet[Theorem 1.11.3]{vw1996}; the second term is nonnegative since the conditional expectation is nonnegative by a generalized Jensen's inequality \citep{ty1975}.
\end{proof}


\section{General Weak Linear IV Models} \label{sec:linear}

This section discusses \cref{exa:iv} where $\pi$ approaches a rank deficient matrix.
We are interested in paths $Q_n$ such that
\begin{gather*}
	\pi(Q_n)=\pi+\frac{\dot{\pi}}{\sqrt{n}}+o\biggl(\frac{1}{\sqrt{n}}\biggr), \qquad
	\beta(Q_n)=\beta+\frac{\dot{\beta}}{\sqrt{n}}+o\biggl(\frac{1}{\sqrt{n}}\biggr), \\
	\gamma(Q_n)=\pi(Q_n)\beta(Q_n)=\gamma+\frac{\dot{\gamma}}{\sqrt{n}}+o\biggl(\frac{1}{\sqrt{n}}\biggr)=\pi\beta+\frac{\dot{\pi}\beta+\pi\dot{\beta}}{\sqrt{n}}+o\biggl(\frac{1}{\sqrt{n}}\biggr),
\end{gather*}
where $\pi$ is of deficient rank $\ell<d$ and $\pi(Q_n)$ is of full column rank for each $n$.

We make use of a few innocuous simplifications to the population model.
First, redefine $z$, $\gamma$, $\pi$ to be $\mathbb{E}_P[zz']^{-1/2}z$, $\mathbb{E}_P[zz']^{1/2}\gamma$, $\mathbb{E}_P[zz']^{1/2}\pi$, so that we have $\mathbb{E}_P[zz']=I$.
Next, by the singular value decomposition, we can write $\pi=USV'$ for a $k\times k$ orthogonal matrix $U$, a $d\times d$ orthogonal matrix $V$, and a $k\times d$ diagonal matrix $S$ whose first $\ell$ elements are positive and all others zero.
Then, by redefining $z$, $x$, $v$, $\gamma$, $\pi$, $\beta$ to be $U'z$, $V'x$, $V'v$, $U'\gamma$, $U'\pi V$, $V'\beta$, we can make $\pi$ equal to $S$.%
\footnote{Note that multiplying an orthogonal matrix to $z$ does not affect $\mathbb{E}_P[zz']=I$.}
To sum up, $\mathbb{E}_P[zz']=I$, $\pi$ is diagonal with its first $\ell$ elements positive, and the last $(\ell-k)\times(\ell-d)$ submatrix of $\dot{\pi}$ is of full column rank.
Henceforth, we adopt the notation:
\[
	\dot{\pi}=\begin{bmatrix}\dot{\pi}_{11}&\dot{\pi}_{12}\\\dot{\pi}_{21}&\dot{\pi}_{22}\end{bmatrix}=\begin{bmatrix}\dot{\pi}_1\\\dot{\pi}_2\end{bmatrix}, \quad
	\beta=\begin{bmatrix}\beta_1\\\beta_2\end{bmatrix}, \quad
	\dot{\beta}=\begin{bmatrix}\dot{\beta}_1\\\dot{\beta}_2\end{bmatrix}, \quad
	\gamma=\begin{bmatrix}\gamma_1\\0\end{bmatrix}, \quad
	\dot{\gamma}=\begin{bmatrix}\dot{\gamma}_1\\\dot{\gamma}_2\end{bmatrix},
\]
where $\pi_{11}$ is an $\ell\times\ell$ matrix, $\pi_1$ is an $\ell\times d$ matrix, and $\beta_1$, $\dot{\beta}_1$, $\gamma_1$, $\dot{\gamma}_1$ are $\ell\times1$ vectors.

We show that $(\gamma,\pi)$ is regular, $\beta$ is weakly regular, and surprisingly, $\beta_1$ is not regular unless $\dot{\pi}_{12}\equiv0$.
Since $\beta_1(Q_n)\to\beta_1(P)=\pi_{11}(P)^\to\gamma_1(P)$, we see that $\beta_1$ is continuous and as such trivially weakly regular.
As before, the score is of the form
\[
	g=g_{uvz}-z'(\dot{\pi}\beta+\pi\dot{\beta})\frac{\frac{\partial}{\partial u}dP_{uvz}}{dP}-z'\dot{\pi}\frac{\frac{\partial}{\partial v}dP_{uvz}}{dP},
\]
and we have
$\mathbb{E}_P[ug\mid z]=z'(\dot{\pi}\beta+\pi\dot{\beta})=z'\dot{\gamma}$ and
$\mathbb{E}_P[v'g\mid z]=z'\dot{\pi}$.
Thus, we find
\[
	\dot{\gamma}_Pg=\mathbb{E}_P[zz']^{-1}\mathbb{E}_P[zug]
	=\begin{bmatrix}\dot{\pi}_{11}\beta_1+\dot{\pi}_{12}\beta_2+\dot{\beta}_1\\\dot{\pi}_{21}\beta_1+\dot{\pi}_{22}\beta_2\end{bmatrix}, \quad
	\dot{\pi}_Pg=\mathbb{E}_P[zz']^{-1}\mathbb{E}_P[zv'g],
\]
showing regularity of $(\gamma,\pi)$.
Moreover, we can rearrange the equality of $\dot{\gamma}_{P,2}$ to write
\[
	\beta_{P,2}(g)=(\dot{\pi}_{P,22}g)^\to(\dot{\gamma}_{P,2}g-(\dot{\pi}_{P,21}g)\beta_1),
\]
which is continuous and homogeneous of degree zero in $g$.
Therefore, $\beta_2$ is weakly regular and so is the entire vector $\beta$.
From the equality of $\dot{\gamma}_{P,1}$,
\[
	\dot{\beta}_{P,1}(g)=\dot{\gamma}_{P,1}g-(\dot{\pi}_{P,11}g)\beta_1-(\dot{\pi}_{P,12}g)\beta_{P,2}(g).
\]
Since $\dot{\gamma}_P$ and $\dot{\pi}_P$ are linear in $g$ and $\beta_{P,2}$ is homogeneous of degree zero in $g$, we see that $\dot{\beta}_{P,1}$ is homogeneous of degree one in $g$.
However, this is not linear in $g$ unless $\dot{\pi}_{P,12}g=0$ for every $g$.
This observation is akin to \Cref{exa:ngmm} where $\pi$ is directionally differentiable but not regular in general.
The expression of $\beta_{P,2}$ indicates that $(\gamma_2,\pi_2)$---not $(\gamma,\pi)$---is a minimal sufficient underlying parameter for $\beta$.

Now we show that 2SLS is a regular estimator.
Since $\mathbb{E}_P[zz']=I$, we can write
\[
	\hat{\beta}_{\text{2SLS}}=(\hat{\pi}'\hat{\pi})^{-1}(\hat{\pi}'\hat{\gamma})+o_P(1)
	=\begin{bsmallmatrix}I\\&\sqrt{n}I\end{bsmallmatrix}\Bigl(\begin{bsmallmatrix}I\\&\sqrt{n}I\end{bsmallmatrix}\hat{\pi}'\hat{\pi}\begin{bsmallmatrix}I\\&\sqrt{n}I\end{bsmallmatrix}\Bigr)^{-1}\begin{bsmallmatrix}I\\&\sqrt{n}I\end{bsmallmatrix}\hat{\pi}'\hat{\gamma}+o_P(1).
\]
Observe that
\begin{multline*}
	\begin{bsmallmatrix}I\\&\sqrt{n}I\end{bsmallmatrix}\hat{\pi}'\hat{\pi}\begin{bsmallmatrix}I\\&\sqrt{n}I\end{bsmallmatrix}
	=\begin{bmatrix}\hat{\pi}_{11}'\hat{\pi}_{11}+\hat{\pi}_{21}'\hat{\pi}_{21}&\sqrt{n}\hat{\pi}_{11}'\hat{\pi}_{12}+\sqrt{n}\hat{\pi}_{21}'\hat{\pi}_{22}\\\sqrt{n}\hat{\pi}_{12}'\hat{\pi}_{11}+\sqrt{n}\hat{\pi}_{22}'\hat{\pi}_{21}&n\hat{\pi}_{12}'\hat{\pi}_{12}+n\hat{\pi}_{22}'\hat{\pi}_{22}\end{bmatrix}\\
	={\underbrace{\begin{bmatrix}\hat{\pi}_{11}'\hat{\pi}_{11}&\sqrt{n}\hat{\pi}_{11}'\hat{\pi}_{12}\\\sqrt{n}\hat{\pi}_{12}'\hat{\pi}_{11}&n\hat{\pi}_{12}'\hat{\pi}_{12}+n\hat{\pi}_{22}'\hat{\pi}_{22}\end{bmatrix}}_{O_P(1)}+\underbrace{\begin{bmatrix}0&\sqrt{n}\hat{\pi}_{21}'\hat{\pi}_{22}\\\sqrt{n}\hat{\pi}_{22}'\hat{\pi}_{21}&0\end{bmatrix}}_{O_P(1/\sqrt{n})}}+o_P\biggl(\frac{1}{\sqrt{n}}\biggr),
\end{multline*}
\[
	\begin{bsmallmatrix}I\\&\sqrt{n}I\end{bsmallmatrix}\hat{\pi}'\hat{\gamma}
	=\begin{bmatrix}\hat{\pi}_{11}'\hat{\gamma}_1+\hat{\pi}_{21}'\hat{\gamma}_2\\\sqrt{n}\hat{\pi}_{12}'\hat{\gamma}_1+\sqrt{n}\hat{\pi}_{22}'\hat{\gamma}_2\end{bmatrix}
	={\underbrace{\begin{bmatrix}\hat{\pi}_{11}'\hat{\gamma}_1\\\sqrt{n}\hat{\pi}_{12}'\hat{\gamma}_1\end{bmatrix}}_{O_P(1)}+\underbrace{\begin{bmatrix}0\\\sqrt{n}\hat{\pi}_{22}'\hat{\gamma}_2\end{bmatrix}}_{O_P(1/\sqrt{n})}}+o_P\biggl(\frac{1}{\sqrt{n}}\biggr).
\]
First, let us focus on the $O_P(1)$ terms.
Write
\[
	H\vcentcolon=\begin{bmatrix}H_{11}&H_{12}\\H_{21}&H_{22}\end{bmatrix}
	\vcentcolon=\begin{bmatrix}\hat{\pi}_{11}'\hat{\pi}_{11}&\sqrt{n}\hat{\pi}_{11}'\hat{\pi}_{12}\\\sqrt{n}\hat{\pi}_{12}'\hat{\pi}_{11}&n\hat{\pi}_{12}'\hat{\pi}_{12}+n\hat{\pi}_{22}'\hat{\pi}_{22}\end{bmatrix}.
\]
By the block matrix inversion formula,
\[
	H^{-1}=\begin{bmatrix}(H_{11}-H_{12}H_{22}^{-1}H_{21})^{-1}&-(H_{11}-H_{12}H_{22}^{-1}H_{21})^{-1}H_{12}H_{22}^{-1}\\-(H_{22}-H_{21}H_{11}^{-1}H_{12})^{-1}H_{21}H_{11}^{-1}&(H_{22}-H_{21}H_{11}^{-1}H_{12})^{-1}\end{bmatrix}.
\]
Thus, the $O_P(1)$ terms make
\[
	H^{-1}\begin{bmatrix}\hat{\pi}_{11}'\hat{\gamma}_1\\\sqrt{n}\hat{\pi}_{12}'\hat{\gamma}_1\end{bmatrix}
	=\begin{bmatrix}(H_{11}-H_{12}H_{22}^{-1}H_{21})^{-1}(\hat{\pi}_{11}'\hat{\gamma}_1-H_{12}H_{22}^{-1}\sqrt{n}\hat{\pi}_{12}'\hat{\gamma}_1)\\(H_{22}-H_{21}H_{11}^{-1}H_{12})^{-1}(\sqrt{n}\hat{\pi}_{12}'\hat{\gamma}_1-H_{21}H_{11}^{-1}\hat{\pi}_{11}'\hat{\gamma}_1)\end{bmatrix}
	=\begin{bmatrix}\hat{\pi}_{11}^{-1}\hat{\gamma}_1\\0\end{bmatrix}.
\]
Therefore, we need the $O_P(1/\sqrt{n})$ terms to derive the asymptotic distribution of $\hat{\beta}_{\text{2SLS},2}$.
They are, by the matrix differentiation formula and the Woodbury matrix identity,
\begin{multline*}
	H^{-1}\begin{bmatrix}0\\\sqrt{n}\hat{\pi}_{22}'\hat{\gamma}_2\end{bmatrix}-H^{-1}\begin{bmatrix}0&\sqrt{n}\hat{\pi}_{21}'\hat{\pi}_{22}\\\sqrt{n}\hat{\pi}_{22}'\hat{\pi}_{21}&0\end{bmatrix}H^{-1}\begin{bmatrix}\hat{\pi}_{11}'\hat{\gamma}_1\\\sqrt{n}\hat{\pi}_{12}'\hat{\gamma}_1\end{bmatrix}\\
	=H^{-1}\begin{bmatrix}0\\\sqrt{n}\hat{\pi}_{22}'(\hat{\gamma}_2-\hat{\pi}_{21}\hat{\pi}_{11}^{-1}\hat{\gamma}_1)\end{bmatrix} 
		=\begin{bmatrix}-\hat{\pi}_{11}^{-1}\sqrt{n}\hat{\pi}_{12}(n\hat{\pi}_{22}'\hat{\pi}_{22})^{-1}\sqrt{n}\hat{\pi}_{22}'(\hat{\gamma}_2-\hat{\pi}_{21}\hat{\pi}_{11}^{-1}\hat{\gamma}_1)\\(n\hat{\pi}_{22}'\hat{\pi}_{22})^{-1}\sqrt{n}\hat{\pi}_{22}'(\hat{\gamma}_2-\hat{\pi}_{21}\hat{\pi}_{11}^{-1}\hat{\gamma}_1)\end{bmatrix}.
\end{multline*}
In short,
\[
	\hat{\beta}_{\text{2SLS}}=\begin{bmatrix}\hat{\pi}_{11}^{-1}\hat{\gamma}_1\\(n\hat{\pi}_{22}'\hat{\pi}_{22})^{-1}\sqrt{n}\hat{\pi}_{22}'(\sqrt{n}\hat{\gamma}_2-\sqrt{n}\hat{\pi}_{21}\hat{\pi}_{11}^{-1}\hat{\gamma}_1)\end{bmatrix}+o_P(1).
\]
Thus, the upper half converges in probability to $\beta_1$ and the lower half to a function of $\sqrt{n}\hat{\gamma}_2$, $\sqrt{n}\hat{\pi}_{21}$, and $\sqrt{n}\hat{\pi}_{22}$, showing regularity of 2SLS.%
\footnote{Note that regularity here is as an estimator of a weakly regular parameter. While $\beta_1$ is weakly regular, it is not regular in general. Therefore, it does not mean that $\sqrt{n}(\hat{\beta}_1-\beta_1)$ is equivariant. In fact, its asymptotic distribution is the limit of $\sqrt{n}(\hat{\pi}_{11}^{-1}\hat{\gamma}_1-\beta_1)-\hat{\pi}_{11}^{-1}\sqrt{n}\hat{\pi}_{12}\hat{\beta}_2$, which is not equivariant even under $\pi_{12}=0$. If we know $\pi_{12}=0$, then running 2SLS with regressors $x_1$ and instruments $z_1$, for example, yields an equivariant estimator for $\beta_1$.}

Under this asymptotics, therefore, we only need to Rao-Blackwellize with respect to $(\hat{\gamma}_2,\hat{\pi}_{21},\hat{\pi}_{22})$.
However, since the coordinate projection of an efficient estimator is efficient and it does not harm to Rao\hyp{}Blackwellize with respect to a strongly identified parameter, we see that the Rao\hyp{}Blackwellized 2SLS for the local\hyp{}to\hyp{}zero asymptotics derived in \cref{exa:iv3} in the main text is also weakly efficient under this asymptotics.


\begin{singlespacing} 
\bibliographystyle{ecta} 
\bibliography{weakbib}
\end{singlespacing}
